\documentclass[aps,prd,twocolumn,amsmath,amsfonts,amssymb,eqsecnum,groupedaddress,nofootinbib,nobalancelastpage,floatfix,superscriptaddress,secnumarabic,notitlepage,10pt]{revtex4-2}
\usepackage[compat=1.1.0]{tikz-feynman}
\usepackage{graphicx, hyperref, multirow, slashed}
\hypersetup{colorlinks=true, linkcolor=blue, citecolor=magenta, filecolor=magenta, urlcolor=cyan}
\usepackage[utf8]{inputenc}
\usepackage[T1]{fontenc}
\DeclareUnicodeCharacter{2212}{-}

\newcommand{\be}{\begin{equation}}
\newcommand{\ee}{\end{equation}}
\def\bsp#1\esp{\begin{split}#1\end{split}}

\begin{document}

\title{Vector Boson Fusion Signatures of Superheavy Majorana Neutrinos at Muon Colliders}

\author{Parham Dehghani}
\email{parham.dehghani@concordia.ca}
\affiliation{Department of Physics, Concordia University, 7141 Sherbrooke St. West, Montreal, Quebec H4B 1R6, Canada}

\author{Mariana Frank}
\email{mariana.frank@concordia.ca}
\affiliation{Department of Physics, Concordia University, 7141 Sherbrooke St. West, Montreal, Quebec H4B 1R6, Canada}

\author{Benjamin Fuks}
\email{fuks@lpthe.jussieu.fr}
\affiliation{Laboratoire de Physique Théorique et Hautes Énergies (LPTHE), UMR 7589, Sorbonne Université et CNRS, 4 place Jussieu, 75252 Paris Cedex 05, France}

\begin{abstract}
We investigate the sensitivity of future high-energy muon colliders to heavy Majorana neutrinos, considering both opposite-sign ($\mu^+\mu^-$) and same-sign ($\mu^+\mu^+$) collision modes. We focus on $\mu^+\mu^-$ colliders operating at centre-of-mass energies of 1, 3 and 10~TeV, as well as the proposed $\mu$TRISTAN facility at 2~TeV, and we analyse the production of heavy neutrinos via vector boson fusion in the $t$-channel, a mechanism that becomes dominant in the multi-TeV regime. We evaluate its exclusion potential in terms of the heavy neutrino mass and the mixing of the heavy neutrino with its Standard Model counterparts, using both cut-based selections and boosted decision trees trained to exploit the distinct kinematic signatures of heavy Majorana neutrino exchanges. Our results  demonstrate the complementarity between collider configurations, and show that active-sterile mixing angles as small as 0.001 could be probed for neutrino masses up to 100~TeV, an experimentally inaccessible region of the parameter space at current facilities. Altogether, this work establishes the discovery potential of muon colliders for testing super-heavy Majorana neutrinos, complementary to conventional probes, and provides compelling motivation for the next generation of high-energy lepton colliders.
\end{abstract}
\maketitle

\section{Introduction}
The discovery of neutrino oscillations stands as one of the most compelling pieces of evidence for physics beyond the Standard Model (BSM). Neutrinos were indeed long believed to be massless, in agreement with the original formulation of the Standard Model (SM) of particle physics. However, experimental anomalies hinted that neutrinos might behave in unexpected ways, which was conclusively confirmed by the Super-Kamiokande~\cite{Super-Kamiokande:1998kpq} and  the Sudbury Neutrino Observatory experiments~\cite{SNO:2002tuh}. These provided clear evidence for neutrino oscillations from one flavour to another, which subsequently implied that neutrinos have non-zero masses and that lepton flavour is not conserved during neutrino propagation in space-time.

These phenomena can be accommodated by introducing fields and interactions supplementing the SM, equipping the theoretical description of nature with a neutrino mass model. The resulting class of BSM setups, collectively named seesaw models~\cite{Minkowski:1977sc, Gell-Mann:1979vob, Glashow:1979nm, Mohapatra:1979ia, Yanagida:1979as, Magg:1980ut, Shrock:1980ct, Schechter:1980gr, Cheng:1980qt, Mohapatra:1980yp, Lazarides:1980nt, Foot:1988aq, Atre:2009rg, Cai:2017mow}, include interactions of new charged or gauge-singlet fermions, scalar states with exotic quantum numbers and/or gauge bosons associated with new symmetries over a large range of masses and couplings. Searches for such signs of physics beyond the SM thus probe one of the most important questions in particle physics today, and may open a window onto new sectors of the BSM theory.

In this work we adopt a model-independent approach and explore BSM setups in which the neutrino sector is extended by one or more heavy neutral leptons, denoted generically as $N$, with masses in the GeV to multi-TeV range. These states are considered to mix with the SM active neutrinos through matrix elements $V_{\ell N}$, and could potentially be observed at present and future high-energy experiments. Here, we focus on the capabilities of proposed future muon-antimuon ($\mu^+\mu^-$) and muon-muon ($\mu^+\mu^+$) colliders designed to probe new physics at the multi-TeV scale, which are expected to offer unique sensitivity to detecting rare processes and heavy states thanks to their clean leptonic environment and high centre-of-mass energy~\cite{Boscolo:2018ytm, Delahaye:2019omf, Long:2020wfp, AlAli:2021let, Black:2022cth, Hamada:2022mua, Accettura:2023ked, Fridell:2023gjx, Akturk:2024evo}.

Several studies have shown that such facilities could significantly increase the sensitivity to heavy neutral leptons and lepton flavour violating processes. For example, in $\mu^+\mu^-$ collisions, a heavy neutrino $N$ can be produced in association with a SM neutrino $\nu$ via $t$-channel $W$-boson or $s$-channel $Z$-boson exchanges. For processes where the heavy neutrino decays into a charged lepton and a $W$ boson through its mixing with the SM, it has been demonstrated that with centre-of-mass energies of 3 to 10~TeV and an integrated luminosity of 1~ab$^{-1}$, searches can probe $N$ masses between approximately 100~GeV and 10~TeV, for mixing angles $|V_{\ell N}|$ in the range of $10^{-2}$ to $10^{-3}$~\cite{Li:2023lkl, Li:2023tbx, Kwok:2023dck}. By comparison, this reach extends significantly beyond that of current and proposed electron-positron and hadron colliders~\cite{Keung:1983uu, delAguila:2007qnc, Banerjee:2015gca, Nemevsek:2018bbt, Pascoli:2018heg, Fuks:2020att, Mekala:2022cmm, Nemevsek:2023hwx, Frank:2023epx, ATLAS:2018dcj, ATLAS:2019isd, ATLAS:2023tkz, ATLAS:2024rzi, CMS:2018agk, CMS:2018jxx, CMS:2021dzb, CMS:2022hvh}.

Complementary searches can be conducted at proposed $\mu^+ e^+$ colliders such as the $\mu$TRISTAN facility. Although limited by a lower centre-of-mass energy of 346~GeV, such setups could still probe mixing angles of order a few times $10^{-3}$ for $N$ masses above approximately 50~GeV and below the production threshold~\cite{Das:2024kyk}. Further opportunities at $\mu$TRISTAN arise from same-sign muon collisions ($\mu^+\mu^+$), which offer two promising production modes~\cite{Jiang:2023mte, deLima:2024ohf}. The first consists of direct $N$ production, which is analogous to the $\mu^+\mu^-$ case and which gives access to heavy neutrino masses extending up to the collider's centre-of-mass energy, expected to be 1 or 2~TeV for mixings of approximately $10^{-3}$. The second mode is an indirect channel where the heavy neutrino mediates a $t$-channel process leading to same-sign $W^+W^+$ pair production. This indirect signature allows sensitivity to $N$ masses ranging up to approximately 1~TeV for small mixings around $|V_{\ell N}| \sim 0.02$, and up to about 100~TeV if the mixing increases to $|V_{\ell N}| \sim 0.2$. In addition, specific bounds have also been derived in complete neutrino mass models, leveraging the presence of extra states at the price of a not so clear interpretation due to a larger number of free parameters~\cite{Dev:2023nha, Bandyopadhyay:2024gyg}.

In this work, we explore an indirect yet powerful probe of heavy neutrinos, focusing both on same-sign and opposite-sign muon pair production via vector boson fusion (VBF) and mediated by $t$-channel exchanges of heavy neutrinos. By construction, this production mechanism, schematically described by the process $\mu^+\mu^\pm \to \nu_\mu\nu_\mu \mu^+\mu^\pm$ (with $\nu_\mu$ standing for a generic neutrino and an antineutrino), has a unique sensitivity to new physics beyond the kinematic threshold of direct production. At the LHC, same-sign dilepton signals via VBF have already been shown to provide constraints on heavy neutrinos with masses well above the collider’s hadronic centre-of-mass energy~\cite{Fuks:2020att, Fuks:2020zbm, CMS:2022hvh, ATLAS:2023tkz, ATLAS:2024rzi}. This enhancement arises from the interplay between the high virtuality of the exchanged gauge bosons and the non-decoupling behaviour of the heavy Majorana neutrino in the $t$-channel. It thus opens a complementary discovery path that is both less reliant on on-shell production and sensitive to the Majorana nature of the heavy states~\cite{Dev:2013wba, Alva:2014gxa}.

We simulate both the signal and the relevant SM backgrounds by means of state-of-the-art event generation, parton showering and detector simulation tools in order to evaluate the projected sensitivity of several benchmark future collider configurations: a $\mu^+\mu^-$ collider operating at a centre-of-mass energy of $\sqrt{s} = 10$~TeV with an integrated luminosity of 10~ab$^{-1}$, as well as lower-energy setups at $\sqrt{s} = 1$ and 3~TeV with 1~ab$^{-1}$. Moreover, we also consider the prospects for a $\mu^+\mu^+$ collider at $\sqrt{s} = 2$~TeV with 1~ab$^{-1}$. Our results demonstrate a remarkable signal and background separation power and that VBF-induced same-sign and opposite-sign muon production allows these machines to probe heavy neutrino masses far beyond the collider energy, reaching up to approximately 100~TeV for active-sterile mixing angles as small as $|V_{\mu N}| \sim 10^{-3}$. This significantly surpasses the reach of other production channels currently studied in the literature, and highlights the exceptional potential of high-energy muon colliders as probes of seesaw-scale physics through indirect production mechanisms.

The remainder of this report is organised as follows. In section~\ref{sec:theo}, we introduce the simplified theoretical framework used to describe generic BSM extensions featuring a heavy neutrino (section~\ref{sec:model}), along with the Monte Carlo simulation toolchain employed in our study (section~\ref{sec:mc}). We also detail our assumptions for the performance of a prospective muon collider detector, and present a dedicated parametrisation tailored to the SFS framework~\cite{Araz:2020lnp, Araz:2021akd, Araz:2023axv} of \textsc{MadAnalysis5}~\cite{Conte:2012fm, Conte:2014zja, Conte:2018vmg}. In section~\ref{sec:pheno}, we discuss our phenomenological analysis, considering both a cut-based selection (section~\ref{sec:cuts}) and a multivariate one (section~\ref{sec:BDT}). We finally summarise and conclude in section~\ref{sec:conclusions}.

\section{Theoretical framework and computational setup}
\label{sec:theo}
\subsection{A simplified model approach for heavy neutrino phenomenology}
\label{sec:model}
To assess the sensitivity of future colliders to heavy neutrinos through indirect VBF probes, we adopt a simplified model inspired by the type-I seesaw mechanism. In this framework, the SM is extended by a single sterile neutrino state $\nu_R$ which mixes with the SM active neutrino gauge eigenstates $\nu_{Le}$, $\nu_{L\mu}$ and $\nu_{L\tau}$~\cite{Atre:2009rg, Cai:2017mow}. The left-handed flavour-eigenstate neutrinos can then be written as
\begin{equation}\label{eq:nu_def}
  \nu_{Li} = \sum_{m=1}^3 U_{im} \nu_m + V_{i N} N\,,
\end{equation}
where $i = e, \mu, \tau$. In this expression, the complex matrix elements $U_{im}$ and $V_{iN}$ parametrise the mixing between the interaction states $\nu_{Li}$ and the mass eigenstates of the light neutrinos $\nu_m$ and the heavy neutrino $N$, respectively. To leading order in the active-sterile mixing elements $V_{\ell N}$, the heavy neutrino therefore inherits interactions with the electroweak sector. These include, in particular, charged-current couplings to the $W$ boson,
\begin{equation}
  \mathcal{L}_{\text{CC}} = 
   - \frac{g}{\sqrt{2}} \sum_{\ell=e,\mu,\tau} \Big( \sum_{i=1}^{3} \overline{\nu}_i\, U_{\ell i}^* +\overline{N} V_{\ell N}^*\Big) \slashed{W} P_L \ell + \mathrm{H.c.}\,,
\end{equation}
where $g$ denotes the weak coupling constant and $P_L$ is the left-handed chirality projector. In the phenomenological approach followed in this work, we focus on the mixing between the heavy neutrino and the muon flavour, \textit{i.e.}\ the mixing matrix element $V_{\mu N}$, which directly controls both the heavy neutrino production rate at muon colliders and its decay width. Rather than imposing constraints from a specific ultraviolet completion or a full seesaw model description, we treat $V_{\mu N}$ and the heavy neutrino mass $M_N$ as free parameters, which allows us to explore the sensitivity of future muon colliders in a model-independent way while retaining the essential features of the seesaw-motivated dynamics.

Within this framework, we consider benchmark points in the parameter space ensuring a width-to-mass ratio $\Gamma_N/M_N$ below 30\% once all two-body decays into SM charged or neutral leptons and electroweak or Higgs bosons are included. This condition ensures that we remain in the regime where the simplified model is expected to yield reliable predictions. In addition, scenarios with $\Gamma_N/M_N \in [10, 30]\%$ will be specifically highlighted, as cross section computations and event generation may suffer from reduced accuracy and require a more careful treatment of the heavy neutrino propagator.

In what follows, the heavy neutrino width is computed with \textsc{MadWidth}~\cite{Alwall:2014bza} and the \textsc{HeavyN}~\cite{Degrande:2016aje} libraries built with \textsc{FeynRules}~\cite{Christensen:2009jx, Alloul:2013bka, Degrande:2014vpa} and then exported to the UFO format~\cite{Degrande:2011ua, Darme:2023jdn}. The obtained predictions show that viable scenarios with neutrino masses above 1~TeV are constrained by the need to limit the value of the mixing matrix element $V_{\mu N}$ since the width scales proportionally to $|V_{\mu N}|^2$. This behaviour is expected, and reflects the enhanced decay rates induced by longitudinally-polarised gauge bosons at high mass~\cite{Atre:2009rg}. For instance, at $M_N = 1$~TeV, mixing values $|V_{\mu N}|^2 \lesssim 0.2$ keep the width-to-mass ratio below 30\%, while at $M_N = 10$~TeV, the same condition requires $|V_{\mu N}|^2 \lesssim 0.002$.

\subsection{Monte Carlo simulations}
\label{sec:mc}
To investigate the considered VBF-induced heavy neutrino signal at future colliders, we employ a state-of-the-art Monte Carlo simulation chain. Hard-scattering signal and background events are generated at leading-order accuracy using the multi-purpose event generator \textsc{MadGraph5\_aMC@NLO}~\cite{Alwall:2014hca}. Whereas the SM implementation is used for the simulation of background events, processes involving heavy neutrinos rely on the \textsc{HeavyN} UFO libraries. The generated parton-level events are subsequently passed to \textsc{Pythia8}~\cite{Bierlich:2022pfr} for parton showering (including QCD and QED radiation) and hadronisation, while detector effects are accounted for using the SFS framework~\cite{Araz:2020lnp, Araz:2021akd, Araz:2023axv} embedded within the \textsc{MadAnalysis5}~\cite{Conte:2012fm, Conte:2014zja, Conte:2018vmg} platform that we also use for the implementation of our analysis. 

For this purpose, we have developed a dedicated SFS detector card that approximates the performance characteristics expected of a future muon collider detector and relies on \textsc{FastJet}~\cite{Cacciari:2008gp, Cacciari:2011ma} for event reconstruction. It incorporates smearing effects that arise predominantly from beam-induced backgrounds originating from in-flight muon decays, which are known to pose a significant challenge to object reconstruction~\cite{Collamati:2021sbv, Aime:2022dnr, MuonCollider:2022ded}. The SFS card also integrates detector performance elements inspired by CLIC studies, such as robust tracking in the presence of beam-induced backgrounds~\cite{CLICdp:2018vnx, Grefe:2012xmo}, as well as design inputs from future circular collider infrastructure which is considered a viable option for muon production and acceleration~\cite{Zimmermann:2018wfu, Benedikt:2022kan}.

\begin{figure*}
  \centering \includegraphics[width=1.0\textwidth]{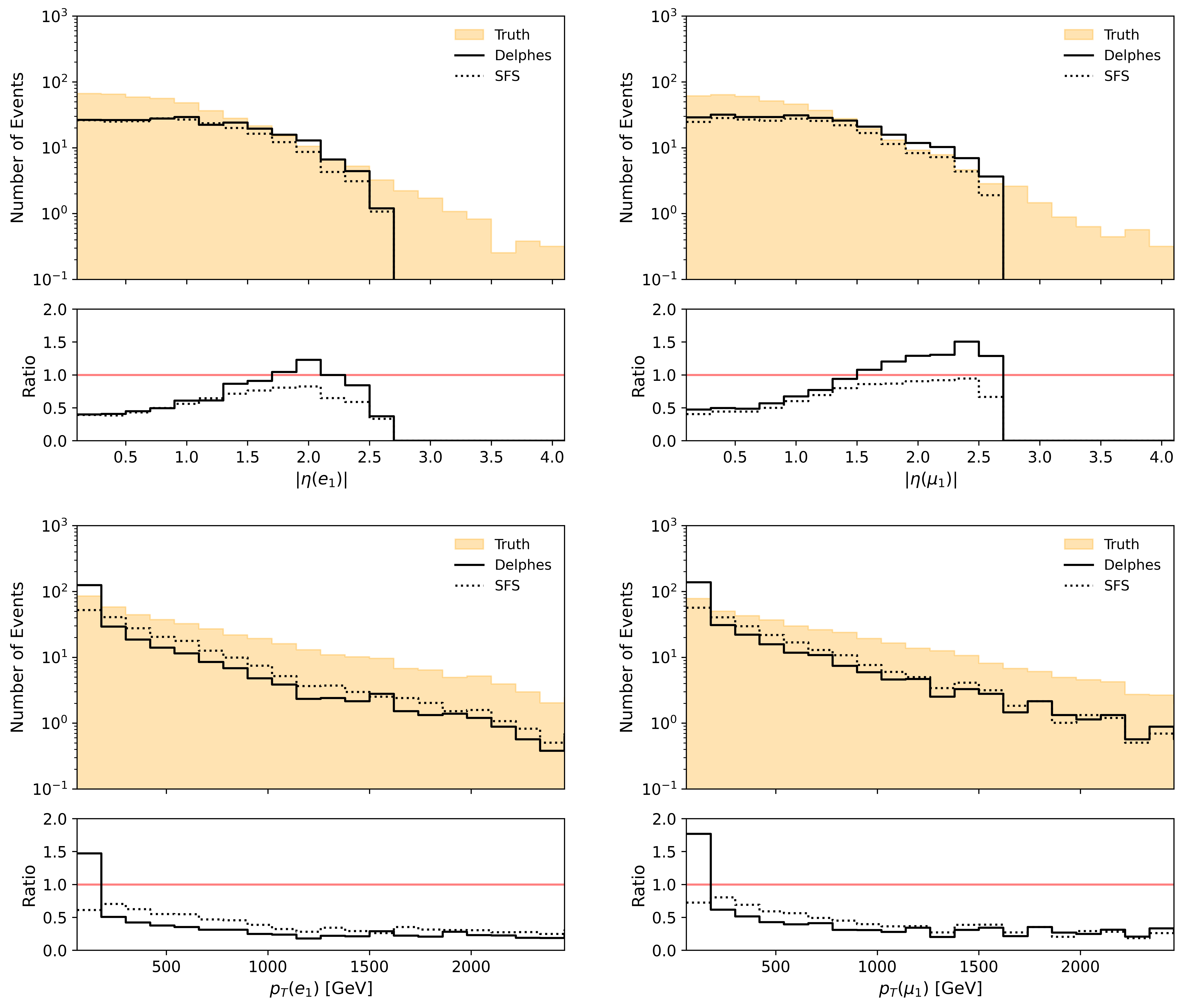}
  \caption{Pseudo-rapidity (top row) and transverse momentum (bottom row) distributions of the leading electron (left) and muon (right). We compare Monte Carlo truth information (shaded orange) to results obtained using \textsc{Delphes} (solid) and our SFS detector parametrisation (dotted). The ratio to the truth distribution is shown in the lower panels of each figure. \label{fig:lep_perf}}
\end{figure*}

We validate our SFS parametrisation using inclusive top-antitop production in a muon collider environment, simulated via the process $\mu^+ \mu^- \to t \bar{t}$ at a centre-of-mass energy of $\sqrt{s} = 10$~TeV. This process is particularly suitable for validation, as it features a diverse set of final-state objects including electrons, muons, taus, photons and jets after parton showering, hadronisation and reconstruction. To emulate realistic detector acceptance, we apply transverse momentum thresholds of 20~GeV for jets and taus, and of 10~GeV for photons and leptons. We then assess the performance of our SFS setup by comparing the reconstructed object properties with those obtained from a more resource-intensive simulation chain based on \textsc{Delphes} version 3.5~\cite{deFavereau:2013fsa} and its muon collider card\footnote{See the repository \url{https://www.github.com/delphes/delphes/blob/master/cards/delphes_card_MuonColliderDet.tcl}.}. In this comparison, any reconstructed object with pseudo-rapidity $|\eta| > 2.5$ is discarded. Additionally, we impose isolation requirements: leptons must be separated from jets by $\Delta R_{\ell j} > 0.1$, otherwise they are removed from the event and considered part of the jet. Isolated photons are similarly required to satisfy $\Delta R_{\gamma j} > 0.2$ with respect to all jets.

In our parametrisation, the reconstruction efficiencies and resolution functions for electrons and muons are derived from the proposed performance of a CLIC-like detector~\cite{CLICdp:2018vnx}. For muons, we implement a constant reconstruction efficiency of 0.999 for energies above 2.5~GeV, and zero otherwise. For electrons, a more fine-grained efficiency function is used, depending on both the electron energy $E$ and pseudo-rapidity $\eta$. For instance, the efficiency is taken to be 0.7 for central and low-energy electrons with \mbox{$3\,\text{GeV} < E < 8\,\text{GeV}$} and \mbox{$0.91 < |\eta| < 1.1$}. The four-momenta of the reconstructed electrons and muons are then smeared using resolution functions that depend on the lepton transverse momentum $p_T$ and pseudo-rapidity $\eta$, and that vary with the lepton energy $E$. As an example, the resolution for a lepton with $2.0 < |\eta| < 2.1$ and $200\,\text{GeV} \leq E < 500\,\text{GeV}$ is set to a constant value of 6.276~MeV. For more details, we refer to the full SFS card implementation, available within \textsc{MadAnalysis5} from version 1.11.1.

The impact of the above performance choices is illustrated in figure~\ref{fig:lep_perf}, where we compare the reconstructed pseudo-rapidity (top row) and transverse momentum (bottom row) distributions of the leading electron (left) and muon (right). We show truth-level predictions (shaded orange), \textsc{Delphes}-based results (solid) and predictions obtained with our SFS implementation (dotted). The lower panels then display the ratio of the reconstructed distributions to the Monte Carlo truth, namely the exact event information available after hadronisation and before the simulation of any detector or reconstruction effect.

Overall, the two detector simulations yield similar lepton spectra, although some important differences emerge due to the distinct approaches to modelling detector effects. However, these differences are consistent with previous findings~\cite{Araz:2020lnp}. In particular, \textsc{Delphes}-based results show a tendency for leptons to migrate toward the forward detector region, leading to an excess of reconstructed leptons compared to the truth-level prediction for pseudo-rapidities $1\lesssim |\eta| \lesssim 2.5$. In contrast, this trend is less pronounced in the SFS simulation, which more closely preserves the shape of the truth-level distribution.

Larger discrepancies are observed in the transverse momentum distributions. Both simulations produce similar high-$p_T$ shapes, the overall efficiency being low due to reduced reconstruction of isolated leptons from boosted top decays. This effect is not surprising and naturally stems from the lepton isolation criteria imposed in the analysis. At low $p_T$, the SFS parametrisation instead better reproduces the truth-level distribution. However, here events simulated using \textsc{Delphes} tends to feature more reconstructed soft leptons, that are simply discarded in the SFS case.

\begin{figure}
  \centering
  \includegraphics[width=.98\columnwidth, trim={0.25cm 0 15.5cm 13cm},clip]{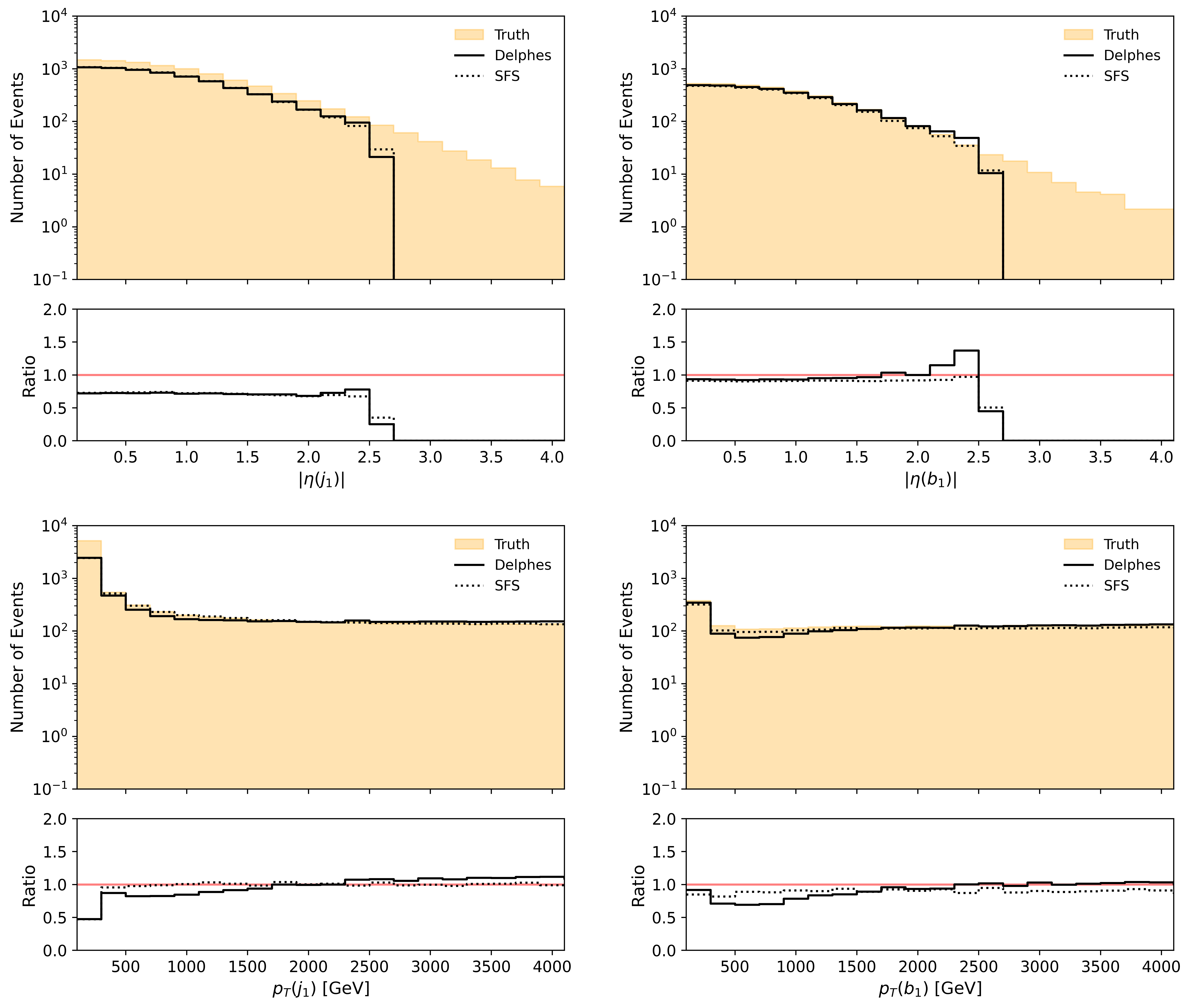}\\ \vspace{.1cm}
  \includegraphics[width=.98\columnwidth, trim={0.25cm 13cm 15.5cm 0},clip]{Figures/figure2.png}
  \caption{Transverse momentum (top) and pseudo-rapidity (bottom) distributions of the leading jet. We compare Monte Carlo truth information (shaded orange) with results from \textsc{Delphes} (solid) and our SFS detector parametrisation (dotted). The lower panels show the ratio to the truth-level distribution.\label{fig:jet_perf}}
\end{figure}

In our parametrisation, jet reconstruction is performed using the anti-$k_T$ algorithm~\cite{Cacciari:2008gp} with a radius parameter $R=0.4$. The impact of the detector includes in particular jet energy resolution effects, that we model through a smearing function inspired by the CLIC detector performance~\cite{CLICdp:2018vnx}, which incorporates noise, stochastic and calorimeter imperfection terms that depend on the jet energy and pseudo-rapidity. For example, in the central region ($|\eta| \leq 0.3$), the energy resolution is modelled as
\begin{equation}
  \frac{\sigma}{E} = \sqrt{\left(\frac{1.38}{E}\right)^2 + \left(\frac{0.308}{\sqrt{E}}\right)^2 + 0.0025}\,,
\end{equation}
which yields an approximate resolution of 5.6\% for a jet with $p_T = 100$~GeV. In addition, we apply an extra transverse momentum smearing of 2\% for central jets ($|\eta| < 0.76$) and of 5\% for more forward jets ($|\eta| \geq 0.76$). 

To assess the impact of this jet modelling, we compare the reconstructed and truth-level distributions of the leading jet’s transverse momentum (top) and pseudo-rapidity (bottom) in figure~\ref{fig:jet_perf}. As before, we contrast results from \textsc{Delphes} (solid) and our SFS implementation (dotted) with the Monte Carlo truth (shaded orange). Both simulation approaches yield remarkably consistent predictions across the full $p_T$ and $|\eta|$ ranges, in contrast with the lepton case discussed earlier where differences were more pronounced due to the treatment of object isolation. We emphasise that comprehensive details can be obtained from the SFS card available within \textsc{MadAnalysis5}.

Additionally, for hadronic tau reconstruction we assume a constant tagging efficiency of 80\% for taus with $p_T \geq 10$~GeV and 0\% otherwise, with light jets being misidentified as hadronic taus with a probability of 2\%. Furthermore, photon reconstruction performance is based on the electromagnetic calorimeter response used for electrons. In the central region of the detector ($|\eta| < 0.7$) the same parametrisation is applied, while slight modifications are implemented in the forward region ($0.7 < |\eta| < 2.5$) as described in~\cite{CLICdp:2018vnx}. This results, for example, in a reconstruction efficiency of 94\% for photons with $E \geq 2$~GeV in the central region, and 90\% in the forward region. The distributions of several properties associated with hadronic taus and photons show very good agreement between the \textsc{Delphes}-based and SFS-based simulations, and are therefore not shown here. As above, we refer to the SFS detector card included with \textsc{MadAnalysis5} version 1.11.1 for more detailed information on the detector modelling. 

\begin{figure}
  \centering
  \includegraphics[width=0.98\columnwidth, trim={0.25cm 0 0.25cm 0.25cm},clip]{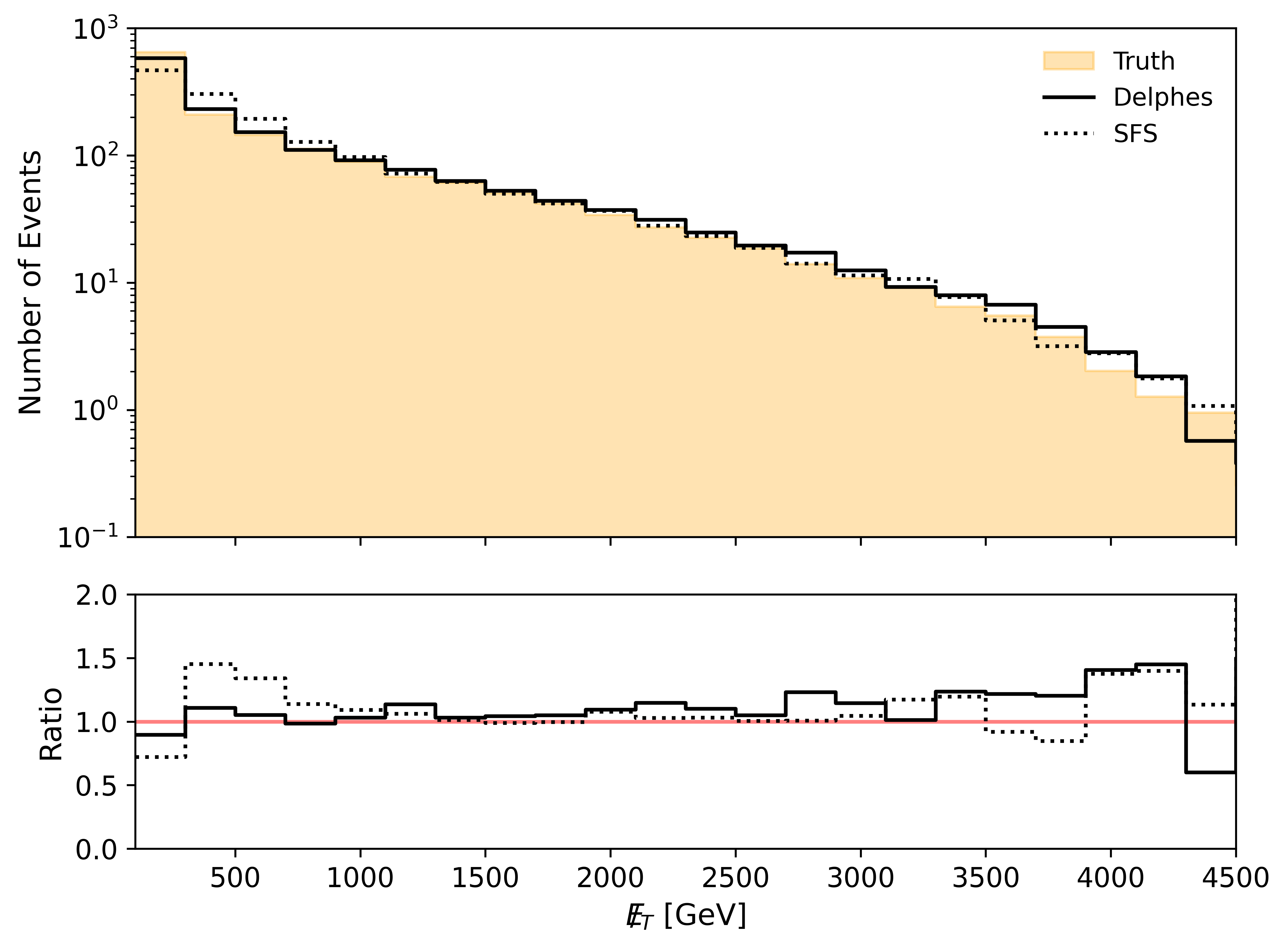}
  \caption{Missing transverse energy distribution derived from Monte Carlo truth information (shaded orange), and from predictions obtained using \textsc{Delphes} (solid) and our SFS detector parametrisation (dotted). The lower panels show the ratio of the reconstructed-level results to the truth-level distribution.\label{fig:met}}
\end{figure}

Before concluding this section, we present in figure~\ref{fig:met} the distribution of the missing transverse energy ($\slashed{E}_T$), which is indirectly sensitive to the reconstruction performance of all visible objects in the event. We again compare Monte Carlo truth information (shaded orange) with the results obtained using \textsc{Delphes} (solid black) and our SFS detector parametrisation (dotted). The figure shows excellent overall agreement between the reconstructed-level and truth-level spectra across most of the $\slashed{E}_T$ range, with the ratio to the Monte Carlo truth remaining close to unity. In the lower part of the spectrum ($\slashed{E}_T \in [200, 800]$~GeV), the SFS simulation predicts slightly more events than \textsc{Delphes}, reflecting differences in the treatment of soft activity. At high $\slashed{E}_T$ values (above 3~TeV), both detector simulations maintain equivalent reconstruction efficiency within the expected statistical uncertainties.

\section{Heavy neutrinos at future muon colliders}
\label{sec:pheno}
\begin{figure}
  \centering
  \begin{tikzpicture}[scale=1.25]\begin{feynman}
    \vertex (i1) at (-1,-.5) ;
    \vertex (i2) at (-1, .5) ;
    \vertex (v1) at (0,0);
    \vertex (v2) at (2,0);
    \vertex (v3) at (3,1);
    \vertex (v4) at (4.5,-.5);
    \vertex (f0) at (3,-1) ;
    \vertex (f1) at (4,2) ;
    \vertex (f2) at (5.5,.5) ;
    \vertex (f3) at (5.5,-1.5) ;
    
    \diagram*{
      (i1) -- [anti fermion, edge label'=\(\mu^+\)] (v1),
      (i2) -- [fermion, edge label=\(\mu^-\)] (v1),
      (v1) -- [boson, edge label=\(Z\)] (v2),
      (v2) -- [plain, edge label=\(N\)] (v3),
      (v2) -- [fermion, edge label=\(\nu_{\mu}\)] (f0),
      (v3) -- [fermion, edge label=\(\nu_{\mu}\)] (f1),
      (v3) -- [boson, edge label=\(Z\)] (v4),
      (v4) -- [anti fermion, edge label=\(\mu^+\)] (f2),
      (v4) -- [fermion, edge label=\(\mu^-\)] (f3),
    };
\end{feynman}\end{tikzpicture}\\ \vspace{.2cm}
  \begin{tikzpicture}[scale=0.75]\begin{feynman}
 \vertex (i1) at (-2,-3) ;
 \vertex (i2) at (-2, 3) ;
    \vertex (v1) at (0,2);  
    \vertex (v2) at (1,1);  
    \vertex (v3) at (1,-1); 
    \vertex (v4) at (0,-2); 
    \vertex (f1) at (3,3) ; 
    \vertex (f2) at (4,1.5) ; 
    \vertex (f3) at (4,-1.5) ; 
    \vertex (f4) at (3,-3) ; 
    
    \diagram*{
      (i1) -- [anti fermion, edge label'=\(\mu^+\)] (v4),
      (i2) -- [fermion, edge label'=\(\mu^-\)] (v1),
      (v1) -- [boson, edge label=\(Z\)] (v2),
      (v2) -- [plain, edge label'=\(N\)] (v3),
      (v3) -- [boson, edge label=\(W^+\)] (v4),
      (v1) -- [fermion, edge label=\(\mu^-\)] (f1),
      (v2) -- [fermion, edge label=\(\nu_{\mu}\)] (f2),
      (v3) -- [anti fermion, edge label=\(\mu^+\)] (f3),
      (v4) -- [anti fermion, edge label'=\(\bar{\nu}_{\mu}\)] (f4),
    };
\end{feynman}\end{tikzpicture}
  \caption{Representative Feynman diagrams for $s$-channel (top) and VBF $t$-channel (bottom) heavy neutrino contribution to the $\mu^+\mu^- \to \mu^+\mu^-\nu_\mu\nu_\mu$ process.\label{fig4:diags}}
\end{figure}
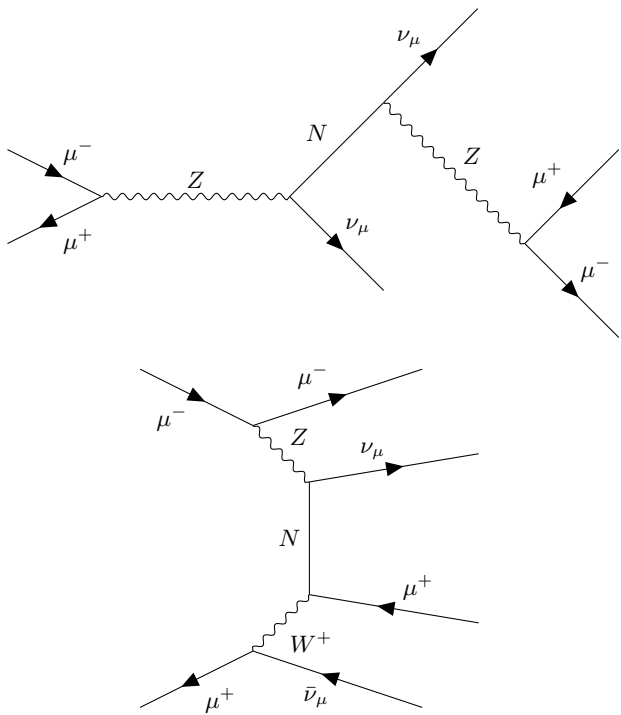

To assess the sensitivity of future muon colliders to heavy Majorana neutrinos produced via vector boson fusion, we consider the processes $\mu^+\mu^- \to \mu^+\mu^-\nu_\mu\nu_\mu$ at a muon-antimuon collider and $\mu^+\mu^+ \to \mu^+\mu^+\nu_\mu\nu_\mu$ at an antimuon-antimuon collider such as the proposed $\mu$TRISTAN machine, where we denote by $\nu_\mu$ both a neutrino and an antineutrino. These 2-to-4 processes, that are potentially lepton-number violating, receive contributions from both $s$-channel and $t$-channel exchanges of heavy Majorana neutrinos, for which we provide representative diagrams in figure~\ref{fig4:diags}. These lead to a cross section scaling in $|V_{\mu N}|^4$ with the relative contributions of the $s$-channel and $t$-channel diagrams which depend strongly on the centre-of-mass energy $\sqrt{s}$ and the heavy neutrino mass $M_N$. The former class of diagrams is dominant for $N$ masses smaller than the centre-of-mass energy, while the latter one only starts to dominate for much larger $M_N$ values.

Our focus is particularly on the $t$-channel VBF topology, which, as shown in the LHC context~\cite{Fuks:2020att, Fuks:2020zbm, CMS:2022hvh, ATLAS:2023tkz, ATLAS:2024rzi}, could provide sensitivity to neutrino masses well above the collider energy. Our study assumes integrated luminosities of 1, 1 and 10 ab$^{-1}$ for $\mu^+\mu^-$ collisions at $\sqrt{s} = 1$, 3 and 10~TeV, respectively, and 1 ab$^{-1}$ for $\mu^+\mu^+$ collisions at $\sqrt{s} = 2$~TeV. The dominant background arises from SM processes that lead to the same final-state topology as well as to a pair of muons, typically mediated by $Z$-bosons, off-shell photons and/or $W$-bosons. For each collider setup, we design an analysis tailored to specific benchmark scenarios and then apply it over a broad range of heavy neutrino masses. Throughout it, we exploit the distinctive kinematic features of the signal to derive exclusion limits in the $(M_N, |V_{\mu N}|)$ parameter space, which we compare to existing bounds from the literature. In section~\ref{sec:cuts}, we present a conventional cut-based analysis employing successive kinematic selections, while section~\ref{sec:BDT} explores a multivariate approach based on boosted decision trees (BDTs), allowing for improved sensitivity in the targeted regions of parameter space.

\subsection{A cut-based analysis of VBF-induced heavy neutrino production at muon colliders}
\label{sec:cuts}

We begin our study with the highest-energy muon collider currently under consideration, operating at a centre-of-mass energy of $\sqrt{s} = 10$~TeV. To establish numerically robust sensitivity projections, we simulate one million signal events across a wide range of heavy neutrino masses spanning $M_N = 10$~GeV to 100~TeV, using the simulation framework described in section~\ref{sec:mc}, together with 10 millions background events. Mixing parameters $V_{\mu N}$ are chosen such that the relative neutrino width $\Gamma_N/M_N$ remains below 30\% throughout.

All signal and background events are subject to baseline object identification criteria: leptons must be isolated from jets with a transverse separation $\Delta R_{\ell j} > 0.4$, and both jets and leptons are required to be central ($|\eta| < 2.5$) and energetic, with minimum transverse momenta of 20~GeV (jets) and 10~GeV (leptons). In particular, the centrality condition plays a crucial role in suppressing large-angle forward muons arising from SM processes such as $\mu^+\mu^- \to \mu^+\mu^-$. To optimise our event selection strategy, we perform a detailed kinematic analysis based on three benchmark scenarios with heavy neutrino masses $M_N = 100$, 500 and 1000~GeV, and respective mixing parameters $V_{\mu N} = 1.0$, 1.0 and 0.5. This set spans a broad region of the parameter space and allows us to design a selection strategy remaining efficient across the full range of interest. The final set of selection criteria is then defined to exploit the distinctive kinematic properties of VBF-induced heavy Majorana neutrino production.

\begin{figure*}
  \centering
  \includegraphics[width=.98\textwidth, trim={0 1.8cm 0 0},clip]{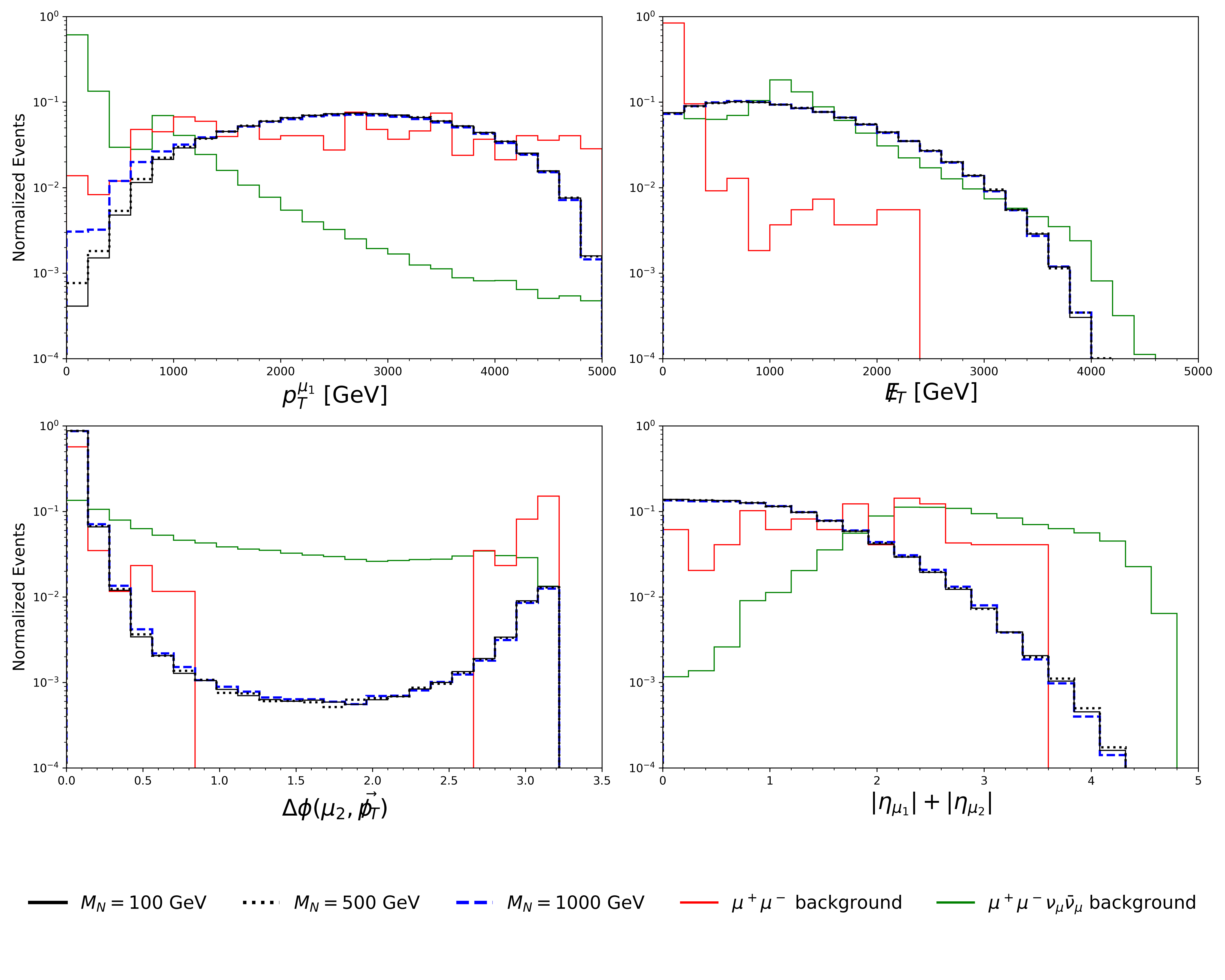}
  \caption{Kinematic distributions used to define the selection cuts shown prior to the application of each respective cut. The panels show distributions for three benchmark signal scenarios corresponding to masses $M_N = 100$, 500 and 1000~GeV and mixing parameters $V_{\mu N} = 1.0$, 1.0 and 0.5, alongside the dominant SM background contributions. We display the transverse momentum of the leading muon (top left), the missing transverse energy (top right), the azimuthal angle between the subleading muon and the missing transverse momentum (bottom left) and the sum of the pseudo-rapidities of the two leading muons (bottom right). \label{fig:kins}}
\end{figure*}

\begin{table*}
  \centering \setlength{\tabcolsep}{8pt} \renewcommand{\arraystretch}{1.3}
  \begin{tabular}{ l | c c c c}
    \multirow{2}{*}{Selection} & Background & Scenario~I & Scenario~II & Scenario~III\\
    & $N_0 = 5.32 \times10^6$& $N_0 = 4.63\times10^9$ & $N_0 = 4.63\times10^9$ & $N_0 = 2.93\times10^8$ \\
    \hline 
    Initial & 1 & 1 & 1 & 1 \\
    2 muons & 0.901 & 0.995 & 0.995 & 0.994 \\
    $M_\mathrm{recoil} < 9500$ GeV & 0.146 & 0.995 & 0.628 & 1.576 \\
    $p^{\mu_1}_T \in [1,4]$~TeV; $p^{\mu_2}_T < 2.4$~TeV & 0.0180 & 0.745 & 0.744 & 0.729 \\
    $\slashed{E}_T > 200$ GeV & 0.0161 & 0.689 & 0.688 & 0.676 \\
    $\Delta\phi(\mu_2, \vec{\slashed{p}}_T) < 0.14$ & 0.00222 & 0.605 & 0.604 & 0.589 \\
    $|\eta_{\mu_1}| + |\eta_{\mu_2}| < 1.0$ & $4.36\times10^{-5}$ & 0.335 & 0.333 & 0.320 \\
    $\eta_{\mu^+} - \eta_{\mu^-} < 0.6$ & $2.33\times10^{-5}$ & 0.273 & 0.271 & 0.261 \\
    $M_{\mu_1\mu_2} > 2500$ GeV & $7.70\times10^{-6}$ & 0.232 & 0.231 & 0.222 \\
  \end{tabular}
  \caption{Cut-flow table showing background and signal efficiencies for three benchmark scenarios (I, II, III) with heavy neutrino masses $M_N = 100$, 500 and 1000~GeV and respective mixing parameters $V_{\mu N} = 1.0$, 1.0 and 0.5, at a $\sqrt{s} = 10$~TeV muon-antimuon collider. Initial event counts $N_0$ are normalised to an integrated luminosity of 10~ab$^{-1}$. \label{tab:cuts}}
\end{table*}

The first selection step requires exactly two reconstructed opposite-sign muons, corresponding to the characteristic final-state topology of the signal. To suppress the SM di-muon background at high recoil energies, we impose a recoil mass constraint of $M_\mathrm{recoil} < 9500$~GeV. Next, we exploit the asymmetric transverse momentum distribution of the signal: the leading muon is required to satisfy $p_T^{\mu_1} \in [1,4]$~TeV while the subleading muon must have $p_T^{\mu_2} < 2.4$~TeV. We then require a minimum missing transverse energy of $\slashed{E}_T > 200$~GeV, consistent with the presence of energetic neutrinos in the final state. A tight azimuthal separation cut $\Delta\phi(\mu_2, \vec{\slashed{p}}_T) < 0.14$ further enhances the signal-to-background ratio by exploiting the strong angular correlation between the subleading muon and the missing momentum, a feature not typically present in SM backgrounds. The top left, top right and bottom left panels of figure~\ref{fig:kins} illustrate the origin and effectiveness of these cuts, comparing signal and background distributions across all three benchmark points.

We further refine the event selection using pseudo-rapidity-based observables. The condition $|\eta_{\mu_1}| + |\eta_{\mu_2}| < 1.0$ ensures that both muons are produced (very) centrally, another feature of the signal illustrated in the bottom right panel of figure~\ref{fig:kins}. In addition, the difference in pseudo-rapidity $\eta_{\mu^+} - \eta_{\mu^-} < 0.6$ serves to suppress backgrounds from processes such as $\mu^+\mu^- \to \mu^+\mu^-\nu_\mu\nu_\mu$, and a final requirement on the di-muon invariant mass $M_{\mu_1\mu_2} > 2500$~GeV is imposed to isolate events consistent with the exchange of a heavy mediator.

The cut efficiencies for both signal and background are summarised in table~\ref{tab:cuts}. For an integrated luminosity of 10~ab$^{-1}$, the background is reduced to approximately 50 events, while retaining substantial signal yields, the exact value depending on $V_{\mu N}$ and $m_N$. The consistent performance of the cuts across different heavy neutrino masses validates the effectiveness of our analysis in capturing the general features of the VBF-induced heavy Majorana neutrino signal regardless of the heavy neutrino mass.

\begin{table}[t]
  \centering \setlength{\tabcolsep}{12pt} \renewcommand{\arraystretch}{1.3}
  \begin{tabular}{l l}
     \multicolumn{2}{c}{Selection}\\
     $\sqrt{s}=$ 3 TeV&$\sqrt{s}=$ 1 TeV \\
     \hline
     2 muons & 2 muons\\
     $M_\mathrm{recoil} < 2400$ GeV &  $M_\mathrm{recoil} < 700$ GeV \\
     $p^{\mu_1}_T < 400$~GeV; $p^{\mu_2}_T < 200$~GeV & - \\
     $\slashed{E}_T > 60$ GeV & $\slashed{E}_T > 30$ GeV\\
     $\Delta\phi(\mu_2,\vec{\slashed{p}}_T) < 1.5$ & $\Delta\phi(\mu_2,\vec{\slashed{p}}_T) > 0.14$\\
     $\eta_{\mu^+} - \eta_{\mu^-} < 2.2$  & $\eta_{\mu^+} - \eta_{\mu^-} < 1.0$\\
     $M_{\mu_1\,\mu_2} > 240$ GeV & $|\eta_{\mu_1}| + |\eta_{\mu_2}| < 1.7$ \\
  \end{tabular} \vspace*{.2cm}
  \begin{tabular}{l}
     Selection at $\sqrt{s} = 2$~TeV \\
     \hline
     2 same-sign muons \\
     $\slashed{E}_T > 40$~GeV \\
     $M_\mathrm{recoil} > 450$~GeV \\
     $p^{\mu_1}_T > 350$~GeV \\
     $\Delta\phi(\mu_2, \vec{\slashed{p}}_T) < 0.4$ \\
     $|\eta_{\mu_1}| + |\eta_{\mu_2}| < 1.7$ \\
     $M_{\mu_1\mu_2} < 1500$~GeV \\
  \end{tabular}
  \caption{Selection criteria targeting the considered heavy neutrino signal at a future muon-antimuon collider expected to operate at $\sqrt{s} = 3$~TeV (top left) and 1~TeV (top right), and at a future same-sign $\mu^+\mu^+$ collider operating at $\sqrt{s} = 2$~TeV (bottom). \label{tab:cutbis}}
\end{table}

Building upon the analysis performed at $\sqrt{s} = 10$~TeV, we extend our study to lower centre-of-mass energies of 3~TeV and 1~TeV, assuming integrated luminosities of 1~ab$^{-1}$ for both scenarios. Due to the modified kinematics and background cross sections at these reduced energies, we design dedicated sets of selection cuts analogous to those considered in the 10~TeV case, targeting the same benchmark scenarios. In particular, at lower centre-of-mass energies, the heavy neutrino signal requires a relaxation of the pseudo-rapidity separation and missing transverse energy criteria compared to the 10~TeV setup. Likewise, the azimuthal correlation between the subleading muon and the missing transverse momentum necessitates a softer requirement. These revised kinematic selection criteria are summarised in the upper panel of table~\ref{tab:cutbis}.

Finally, we investigate the prospects for heavy Majorana neutrino production via vector boson fusion at the $\mu$TRISTAN collider, focusing on the same-sign $\mu^+\mu^+$ initial state. In our analysis, we consider an integrated luminosity of 1~ab$^{-1}$ and a centre-of-mass energy of 2~TeV, consistent with the $\mu$TRISTAN design parameters~\cite{Hamada:2022mua}. We target the process $\mu^+ \mu^+ \to \mu^+ \mu^+ \nu_\mu \nu_\mu$, where we denote both neutrinos and antineutrinos generically by $\nu_\mu$, like for the opposite-sign collider case. Notably, the same-sign configuration only allows for the VBF $t$-channel topology, excluding $s$-channel diagrams that contribute in the opposite-sign scenario.

\begin{figure*}
  \centering \includegraphics[width=0.97\textwidth, trim={0 2cm 0 0},clip]{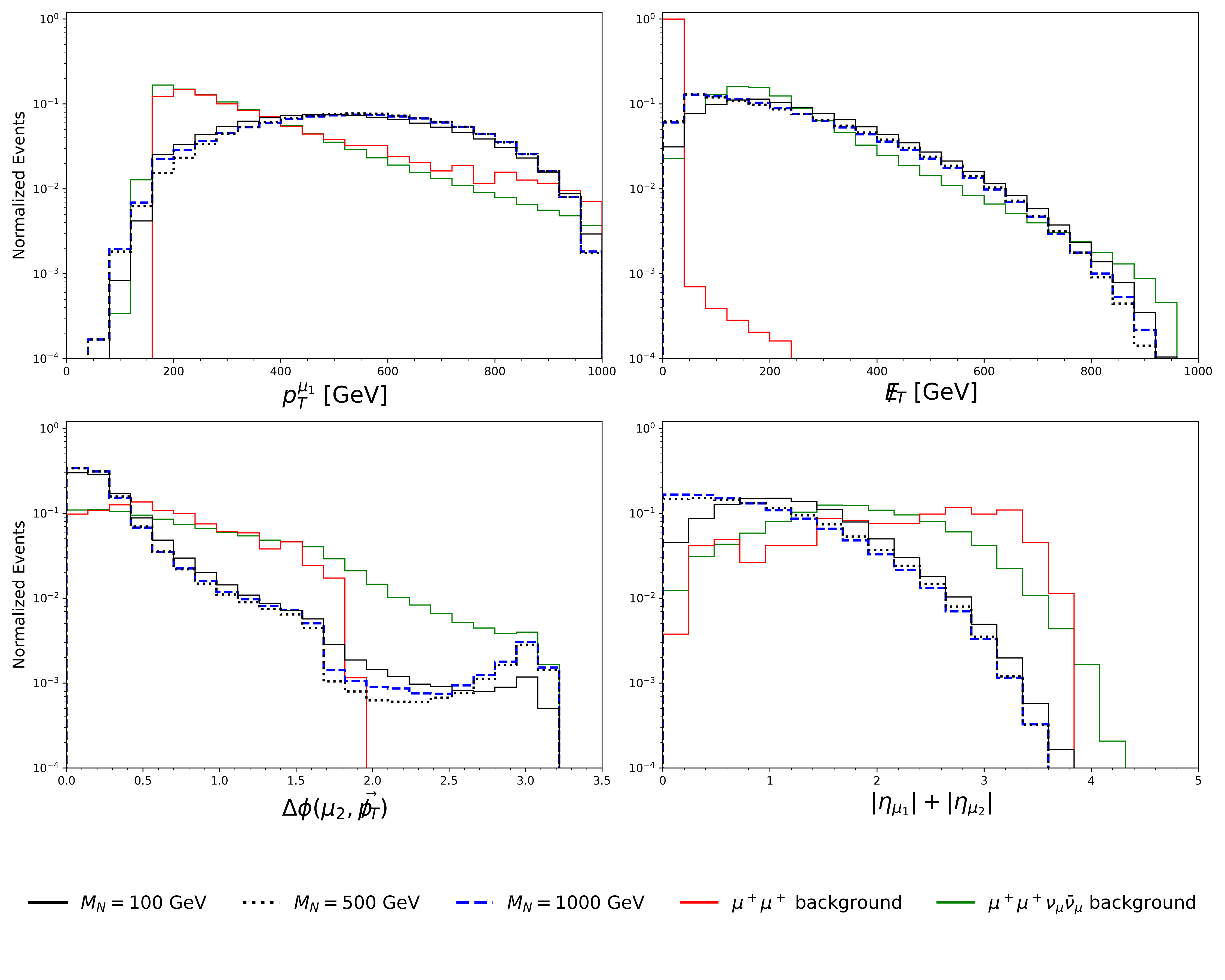}
  \caption{Same as figure~\ref{fig:kins} but for the $\mu$TRISTAN collider configuration. \label{fig:kinsbis}}\vspace{.25cm}
  \includegraphics[width=0.97\textwidth]{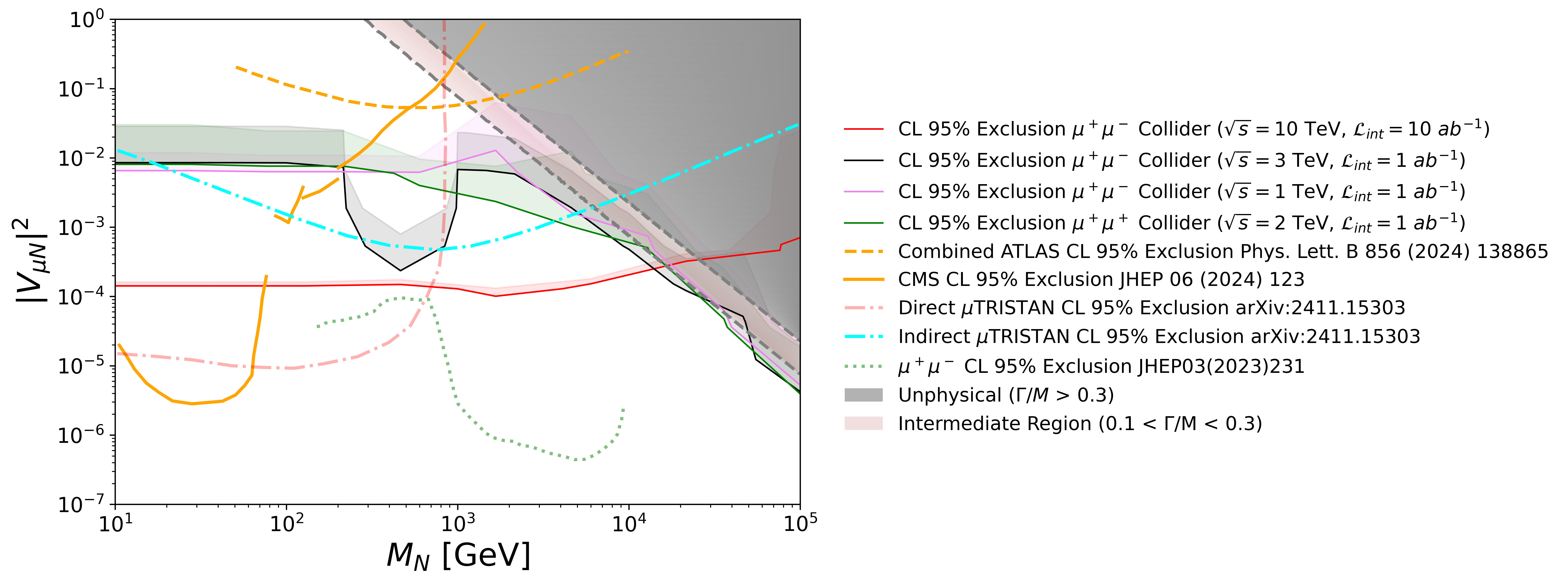}
  \caption{95\% CL exclusions expected from the four muon colliders considered, projected in the $(M_N, |V_{\mu N}|^2)$ plane. The red, blue and purple contours correspond to 10, 1 and 1 ab$^{-1}$  of $\mu^+\mu^-$ collisions at $\sqrt{s} = 10$, 3 and 1~TeV, respectively while the green contours correspond 1 ab$^{-1}$ of $\mu^+\mu^+$ collisions at $\sqrt{s} = 2$~TeV, with band thickness reflecting background systematic uncertainties varying from 0\% to 20\%. In addition, we display the most up-to-date limits from the LHC~\cite{ATLAS:2024rzi, CMS:2024xdq} (dashed and solid orange) and the $\mu$TRISTAN projections of~\cite{deLima:2024ohf} for direct (dash-dotted pink) and indirect (dash-dotted teal) searches for heavy neutrinos, as well as the projections from~\cite{Li:2023tbx} for conventional searches at $\mu^+\mu^-$ colliders (dotted green). Unphysical regions where $\Gamma_N/M_N > 0.3$ (dark grey) and intermediate regions where $0.1 < \Gamma_N/M_N < 0.3$ (light pink) are also shown. \label{fig:res_cutcount}}
\end{figure*}

Our analysis strategy mirrors the one developed for the considered muon-antimuon collider cases, with optimised cuts tailored to three benchmark scenarios featuring heavy neutrino masses of $M_N = 100$, 500 and 1000~GeV. The selection criteria are summarised in the lower panel of table~\ref{tab:cutbis} and representative kinematic distributions are shown in figure~\ref{fig:kinsbis}. Events are selected based on the presence of significant recoil energy carried by the invisible neutrinos, contrasting with the opposite-sign case where limiting this recoil helped to suppress the SM background. This reflects the dominant $t$-channel nature of the signal in the same-sign configuration, and for the same reason, a stringent cut is also applied on the transverse momentum of the leading muon, $p_T^{\mu_1} > 350$~GeV. Furthermore, the azimuthal correlation between the subleading muon and the missing transverse momentum displays a distinctive behaviour, which we exploit to enhance signal sensitivity while significantly reducing background contamination. Additional discriminating power is finally achieved using the pseudo-rapidity and di-muon invariant mass distributions.

In order to evaluate the parameter space coverage offered by our various analyses, we select benchmark points for each collider configuration spanning the relevant heavy neutrino mass range while ensuring that the width-to-mass ratio satisfies $\Gamma_N/M_N < 30\%$. For each benchmark, we compute the signal significance $\sigma$ using the standard expression
\begin{equation}
  \sigma = \frac{S}{\sqrt{S + B + \sigma_B^2}}\,,
\end{equation}
where $S$ and $B$ denote the number of signal and background events passing the entire selection, respectively, and $\sigma_B^2$ represents the background uncertainty which we vary from an idealised 0\% to a conservative 20\% of the background yield. Leveraging the quartic dependence of the signal cross section on $|V_{\mu N}|$, we then derive 95\% confidence level (CL) exclusion limits in the $(M_N, |V_{\mu N}|^2)$ plane, as shown in figure~\ref{fig:res_cutcount}. We present results for the four considered muon collider configurations: $\mu^+\mu^-$ collisions at $\sqrt{s} = 10$~TeV (red), 3~TeV (blue) and 1~TeV (purple) assuming integrated luminosities of 10~ab$^{-1}$, 1~ab$^{-1}$ and 1~ab$^{-1}$ respectively, as well as $\mu^+\mu^+$ collisions at a $\sqrt{s} = 2$~TeV $\mu$TRISTAN (green) with a luminosity of 1~ab$^{-1}$. For each curve, the thickness of the exclusion contour reflects the variation in systematic background uncertainties. 

Our results indicate that sensitivity to heavy neutrinos with masses up to 100~TeV is technically possible, but only in extreme cases featuring $\Gamma_N/M_N > 30\%$. More realistic scenarios with $\Gamma_N/M_N < 30\%$ remain accessible up to around 20~TeV, provided the active-sterile mixing is of order a few percent. Notably, all collider configurations demonstrate sensitivity in the very-heavy regime thanks to their differing background characteristics and the dedicated selection strategies that we have designed. Their impact becomes evident when comparing our limits not only with those from~\cite{deLima:2024ohf} (dash-dotted curves) and~\cite{Li:2023tbx} (dotted contour), both of which relying on conventional search channels, but also with those of the LHC~\cite{ATLAS:2024rzi, CMS:2024xdq} (orange). We find that for each setup, the VBF-induced production of heavy neutrinos provides a complementary probe to conventional direct and indirect searches, enabling sensitivity especially when $M_N \gg \sqrt{s}$. This complementarity is particularly evident for $\mu$TRISTAN, where meaningful constraints are even achievable down to $M_N \sim 500$~GeV. Conversely, for masses below the centre-of-mass energy, standard search strategies remain more effective, often outperforming the VBF-based limits by up to an order of magnitude depending on the heavy neutrino mass and collider configuration. 

An important feature of our results is the change in the shape of the exclusion contours around $M_N \approx 0.5-1$~TeV. This transition marks the shift in the dominant production mode from $s$-channel diagrams to $t$-channel VBF topologies, which is particularly visible from the 3~TeV collider exclusion contour. Here, for $M_N$ values below 500~GeV, $s$-channel production maintains favourable signal rates while for $M_N$ values higher than 1~TeV, the signal turns out to be dominated by VBF production. In the transition region, interference between the two channels finally yields an enhancement of the production cross section, implying a much more significant sensitivity. Such a behaviour is also seen in the 1~TeV case, but in a more minimal way, while for the other setups there is no sizeable $s$-channel contribution. Beyond the 1~TeV mass threshold, VBF production ensures that the cross section remains sizeable, even up to $M_N \sim 30$~TeV. Interestingly, above about 10~TeV, lower-energy muon colliders can surpass the sensitivity of a 10~TeV muon collider, which underscores the superior efficiency of VBF mechanisms at high masses. It also highlights the non-trivial interplay between collider energy and production topology: a higher centre-of-mass energy does not always guarantee better sensitivity. While the signal cross section decreases with increasing heavy neutrino mass, scaling as $|V_{\mu N}|^4/M_N^4$ for large $M_N$ values~\cite{Fuks:2020att}, the interplay between the collider energy and the VBF topology exploited by our selection cuts leads to enhanced sensitivity in this regime. The strong background suppression achieved by the analysis ensures that even a small number of signal events can yield significant sensitivity. This effect is especially pronounced at lower collision energies, where the SM backgrounds are kinematically suppressed due to the absence of any dominant VBF contributions, unlike at $\sqrt{s} = 10$~TeV where the remaining background resembles the signal topology. As a result, for heavy neutrino masses above 10~TeV, lower-energy muon colliders could be expected to outperform a 10~TeV collider in terms of sensitivity. Finally, we note that our results remain robust against the inclusion of background systematic uncertainties up to 20\%, with only minor degradation in sensitivity observed across the explored parameter space.

\subsection{Improving sensitivity to VBF-induced heavy neutrino production with BDTs} 
\label{sec:BDT}
To refine our assessment of the reach of future muon colliders to heavy Majorana neutrinos, we rely on gradient boosting techniques to improve the 95\% confidence level exclusions obtained from the traditional cut-based analyses implemented in section~\ref{sec:cuts}. This approach allows us to exploit the full multidimensional space of observables used so far without the need for manual selections. Unlike simple sequential cut-based methods, boosted decision trees (BDTs) can hence capture non-linear correlations between variables and better exploit the distinctive topological characteristics of a signal~\cite{Roe:2004na, Coadou:2013lca, Albertsson:2018maf, Carleo:2019ptp, Cornell:2021gut, Choudhury:2024crp, Cornell:2024dki}. To quantify the potential improvements afforded by this technique in the model under study, we focus on two benchmark colliders: the most optimistic muon collider setup that we have considered, namely $\mu^+\mu^-$ collisions at $\sqrt{s} = 10$~TeV with a luminosity of 10~ab$^{-1}$, and the $\mu$TRISTAN configuration corresponding to $\mu^+\mu^+$ collisions at $\sqrt{s} = 2$~TeV with a luminosity of 1~ab$^{-1}$. In both cases, we specifically target VBF-mediated heavy neutrino indirect production at very large masses.

\begin{table*}
  \centering \setlength{\tabcolsep}{10pt} \renewcommand{\arraystretch}{1.3}
  \begin{tabular}{c|c c c|c c c}
    \multirow{2}{*}{Hyperparameter}& \multicolumn{3}{c|}{Opposite-sign} & \multicolumn{3}{c}{Same-sign} \\
     & I & II & III & IV & V & VI \\
    \hline
    \texttt{max\_depth} & 10 & 3 & 3 & 3 & 10 & 10 \\
    \texttt{learning\_rate} & 0.05 & 0.05 & 0.1 & 0.05 & 0.05 & 0.1 \\
    \texttt{n\_estimators} & 200 & 500 & 300 & 500 & 300 & 400 \\
    \texttt{min\_child\_weight} & 5 & 1 & 3 & 1 & 3 & 2 \\
    \texttt{gamma} & 0.1 & 0.1 & 0.2 & 0.1 & 0.2 & 0.2 \\
    \texttt{subsample} & 0.80 & 0.95 & 0.95 & 0.95 & 0.80 & 0.80 \\
    \texttt{colsample\_bytree} & 0.8 & 0.8 & 0.9 & 0.8 & 0.8 & 0.8 \\
    \texttt{alpha} & 0.4 & 0.2 & 0.4 & 0.2 & 0.2 & 0.2 \\
    \texttt{early\_stopping\_rounds} & 20 & 20 & 20 & 20 & 20 & 20 \\
  \end{tabular}
  \caption{Optimised hyperparameters for BDT models trained on three heavy Majorana neutrino benchmark scenarios for 10~TeV opposite-sign and 2~TeV same-sign muon collider configurations. The scenarios I, II, III, IV, V and VI are defined by $(M_N, V_{\mu N}) = (100~\mathrm{GeV}, 1.0)$, $(5~\mathrm{TeV}, 0.1)$, $(20~\mathrm{TeV}, 0.025)$, $(500~\mathrm{GeV}, 0.9)$, $(2~\mathrm{TeV}, 0.2)$ and $(10~\mathrm{TeV}, 0.05)$ respectively, and all models are trained using binary logistic regression~\cite{Hastie:2009itz} with the area-under-curve (AUC) as the evaluation metric.\label{tab:hyperprms}}
\end{table*}

For the $\mu^+\mu^-$ collider configuration, we consider three benchmark scenarios with $M_N = 100$~GeV, 5~TeV and 20~TeV and corresponding active-sterile mixing parameters set to $V_{\mu N} = 1.0$, $0.1$ and $0.025$, respectively. For the same-sign $\mu^+\mu^+$ case, we instead choose $M_N = 500$~GeV, 2~TeV and 10~TeV. For both signal and background samples, we apply the same baseline object selection criteria as defined in section~\ref{sec:cuts}. In particular, leptons are required to be isolated from jets by $\Delta R_{\ell j} > 0.4$, and all jets and leptons must satisfy $|\eta| < 2.5$ with $p_T > 20$~GeV for jets and $p_T > 10$~GeV for leptons. To train the BDT classifier, we construct a comprehensive set of kinematic features for each event. These include the transverse momentum, pseudo-rapidity and azimuthal angle of the leading and subleading muons ($p_T^{\mu_1}$, $p_T^{\mu_2}$, $\eta_{\mu_1}$, $\eta_{\mu_2}$, $\phi_{\mu_1}$, $\phi_{\mu_2}$), as well as properties of the di-muon system such as its invariant mass $M_{\mu_1\mu_2}$, the azimuthal separation $\Delta\phi(\mu_1, \mu_2)$  and the recoil mass $M_\mathrm{recoil}$. We further include derived quantities such as the transverse momentum ratio $p_T^{\mu_1}/p_T^{\mu_2}$, the sum $p_T^{\mu_1}+p_T^{\mu_2}$, the pseudo-rapidity sum $|\eta_{\mu_1}| + |\eta_{\mu_2}|$, the difference $\eta_{\mu^+} - \eta_{\mu^-}$ and the angular distance $\Delta R(\mu_1, \mu_2)$. In addition, we consider the missing transverse energy $\slashed{E}_T$, its angular separation from each muon $\Delta\phi(\mu_1, \vec{\slashed{p}}_T)$ and $\Delta\phi(\mu_2, \vec{\slashed{p}}_T)$, and the ratio $\slashed{E}_T / (p_T^{\mu_1} + p_T^{\mu_2})$. Although several of these variables are correlated, we will show below that only a reduced subset of them significantly drives the sensitivity of the analysis.

Our BDT analysis is implemented using the \textsc{XGBoost} package~\cite{Chen:2016btl}, which offers several advantages for high-energy physics applications. These include robust regularisation, efficient handling of sparse data and parallel tree construction. The signal and background samples are split into training (70\%) and test (30\%) sets, using stratified sampling to preserve the signal-to-background ratio across subsets. To prevent overfitting, we further split the training data into a training subset (80\%) and a validation subset (20\%), enabling early stopping during training when the validation loss ceases to improve. Finally, the BDT hyperparameters are optimised to maximise the area under the receiver operating characteristic (ROC) curve for each of the considered signal scenarios. 

\begin{figure*}
  \centering
  \includegraphics[width=0.495\textwidth]{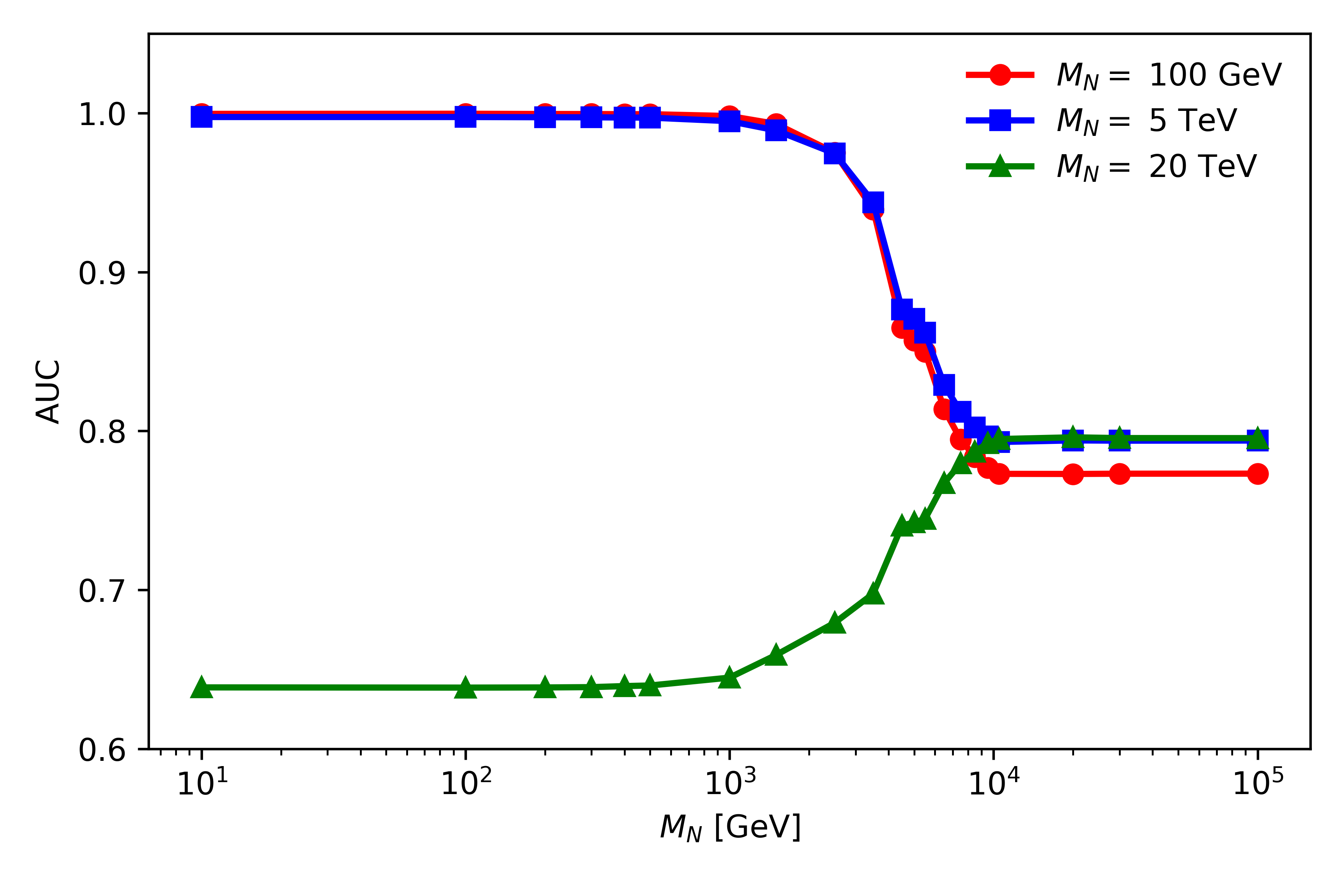}\hfill
  \includegraphics[width=0.495\textwidth]{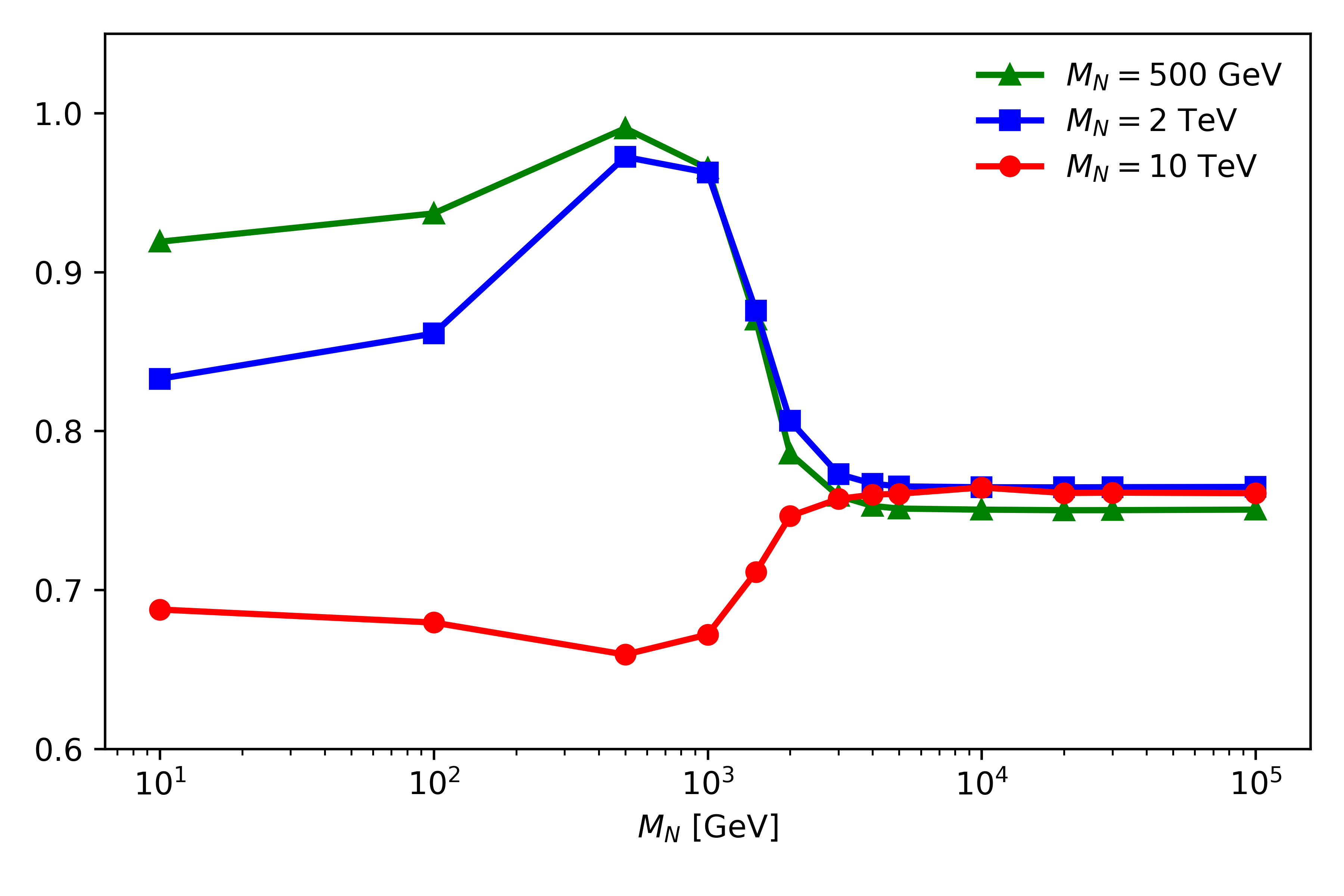}
  \caption{Classification performance of BDT models trained for three representative heavy Majorana neutrino benchmarks with masses $M_N$ fixed to 100~GeV, 5~TeV and 20~TeV for $\mu^+\mu^-$ collisions at $\sqrt{s}=10$~TeV (left) and 500~GeV, 2~TeV and 10~TeV for $\mu^+\mu^+$ collisions at $\sqrt{s}=2$~TeV (right). The performance is quantified using the area under the ROC curve, plotted as a function of the heavy neutrino mass $M_N$. \label{fig:auc}}
\end{figure*}

To illustrate our findings, we summarise the resulting BDT configurations for each benchmark scenario in table~\ref{tab:hyperprms}, for the case of 10~TeV muon-antimuon collisions (second, third and fourth columns) and $\mu^+\mu^+$ collisions at 2~TeV (last three columns). For the 10~TeV $\mu^+\mu^-$ configuration, the hyperparameter choices vary across benchmarks, reflecting different classification challenges as the production mechanism shifts from mixed-channel to VBF-dominated regimes. For instance, the 100~GeV model uses significantly deeper trees (a maximum depth of 10), suggesting the need for more complex decision boundaries when both $s$-channel and $t$-channel contributions are relevant. Meanwhile, the number of estimators is largest in the intermediate 5~TeV mass scenario (500 estimators), which suggests that model complexity peaks in the intermediate mass regime. For the highest mass scenario, the learning rate increases to 0.1 and the tree feature sampling rises to 0.9, these adjustments improving the training efficiency and stability of gradient descent in the presence of the large-momentum-transfer signatures typical of VBF-driven processes. The resulting BDT classifiers achieve area-under-curve (AUC) values of 0.999, 0.869, and 0.795 for the 100~GeV, 5~TeV and 20~TeV benchmarks, respectively, demonstrating strong classification performance across a broad range of signal topologies. In contrast, for the 2~TeV $\mu^+\mu^+$ collider configuration, the hyperparameters are more similar across the three benchmark scenarios as the signal topology is more uniform. However, comparing the results in the last three columns of table~\ref{tab:hyperprms} highlights the more complex situation corresponding to the lightest scenario, yielding a signal harder to distinguish from the background. This is handled by requiring less deep trees with more estimators.

For simplicity we select a single BDT model that can be used uniformly across the entire heavy neutrino mass range. We evaluate the classification performance of the three trained models on test samples spanning $M_N$ values from 10~GeV to 100~TeV. Figure~\ref{fig:auc} shows the AUC as a function of $M_N$ for $\mu^+\mu^-$ collisions at $\sqrt{s}=10$~TeV (left) and $\mu^+\mu^+$ collisions at $\sqrt{s}=2$~TeV (right). For the case of opposite-sign muon collisions, the best-performing model across most of the mass range is the one trained at $M_N = 5$~TeV. While the model trained at 100~GeV excels in the low-mass region ($M_N \lesssim 1$~TeV), its performance drops significantly for heavier neutrino masses where VBF contributions dominate. Conversely, the 20~TeV-trained model captures the high-mass behaviour reasonably well but performs poorly in the low-mass regime. The 5~TeV model subsequently offers a balanced performance across the entire mass range and successfully tracks the transition from $s$-channel to VBF topologies. It thus makes it a suitable compromise for a global analysis. 

For same-sign muon collisions at $\sqrt{s}=2$~TeV, the signal is VBF-dominated across the full mass range. Interestingly, in this case, the low-mass (500~GeV) model performs slightly better than the others, likely due to its training on VBF-like topologies already prominent at low $M_N$ values requiring a different BDT model. Based on these observations, we adopt the 5~TeV BDT model for $\mu^+\mu^-$ collisions and the 500~GeV model for $\mu^+\mu^+$ collisions.

\begin{figure*}
  \centering
 \includegraphics[width=0.495\textwidth]{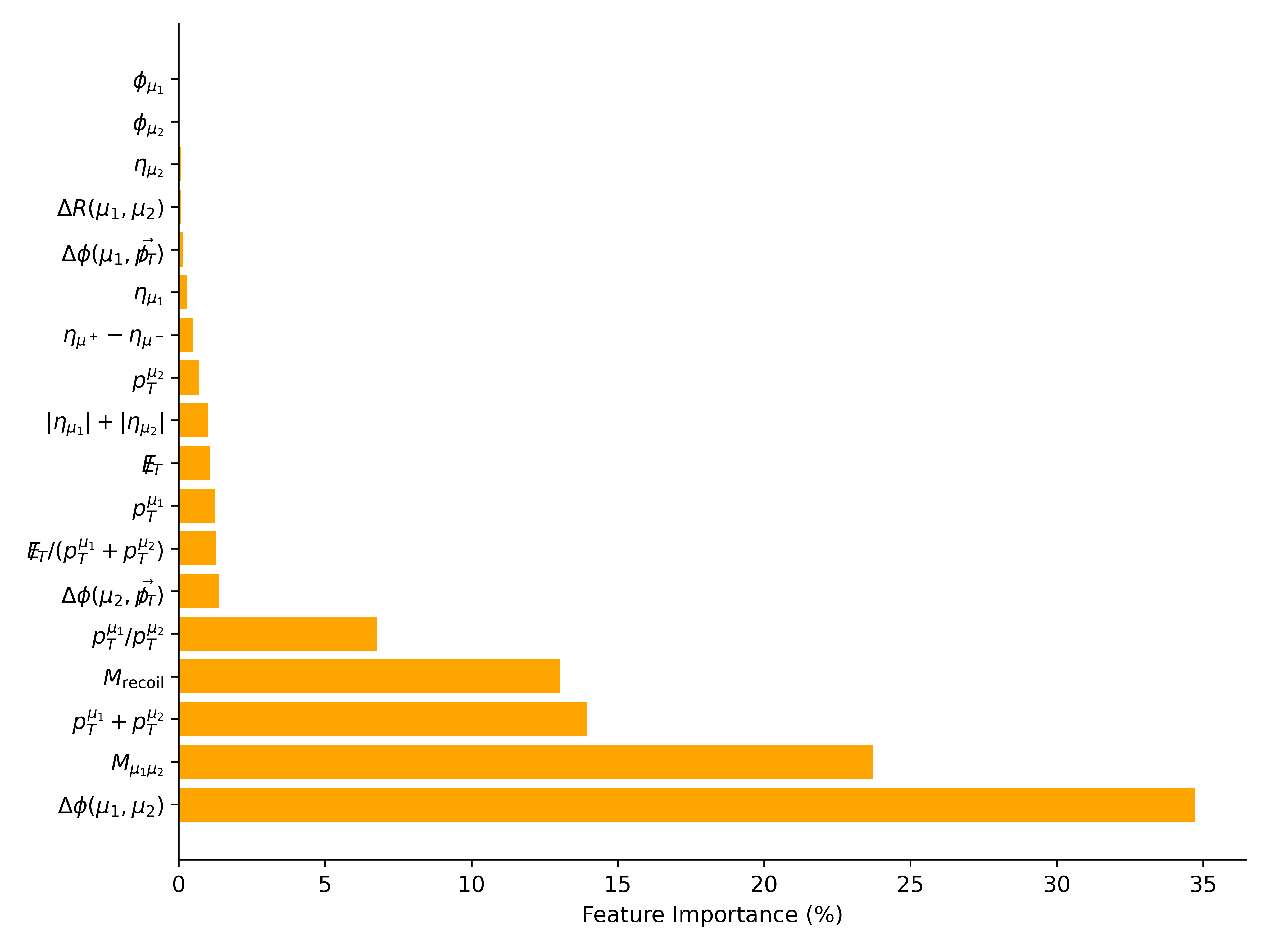}\hfill
  \includegraphics[width=0.495\textwidth]{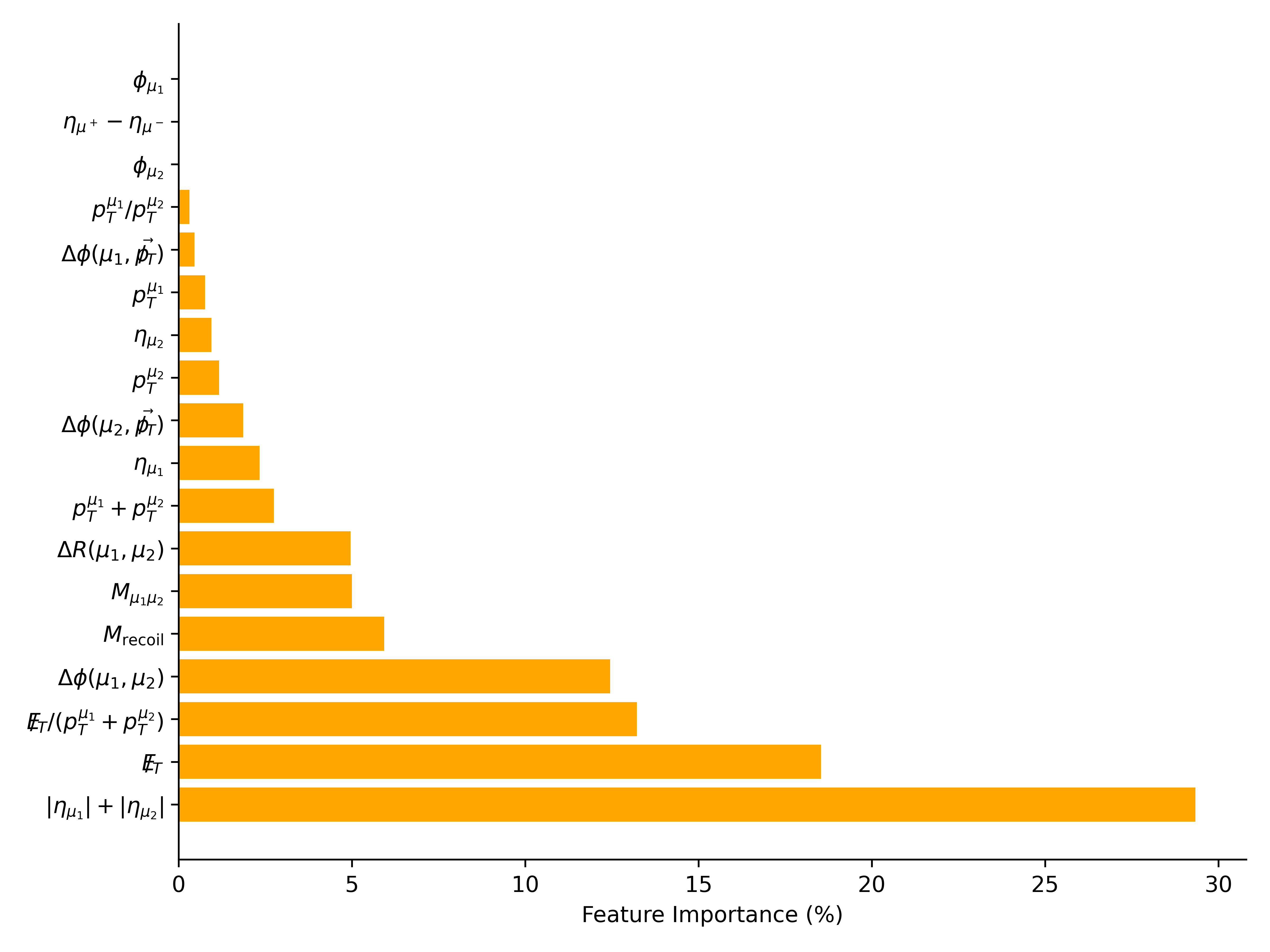}
  \caption{Importance ranking calculated using the gain-based method implemented in \textsc{XGBoost}, for the opposite-sign muon collider configuration at $\sqrt{s}=10$~TeV (left) and the same-sign muon collider configuration at $\sqrt{s}=2$~TeV (right).\label{fig:importance}}
\end{figure*}

\begin{figure*}
  \centering
  \includegraphics[width=.98\textwidth]{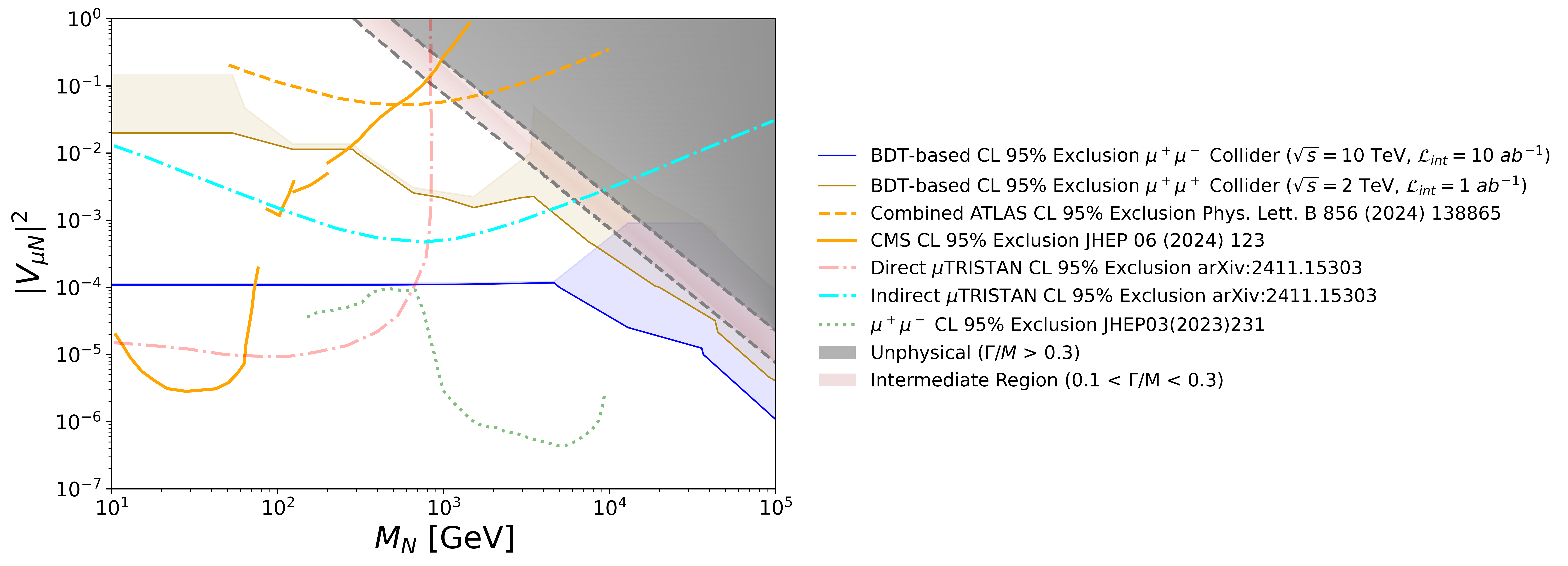}
  \caption{Projected 95\% CL exclusion limits in the $(M_N, |V_{\mu N}|^2)$ plane using our two BDT-based analyses. The blue contour corresponds to $\mu^+\mu^-$ collisions at $\sqrt{s} = 10$~TeV with an integrated luminosity of 10~ab$^{-1}$, while the olive contour shows results for $\mu^+\mu^+$ collisions at $\sqrt{s} = 2$~TeV with 1~ab$^{-1}$. The thickness of the band reflects the impact of background systematic uncertainties, ranging from 0\% to 20\%. For comparison, existing LHC limits~\cite{ATLAS:2024rzi, CMS:2024xdq} (dashed and solid orange) are shown, along with direct (dash-dotted pink) and indirect (dash-dotted teal) projections from $\mu$TRISTAN~\cite{deLima:2024ohf} and estimates from conventional $\mu^+\mu^-$ searches at future colliders~\cite{Li:2023tbx} (dotted green). Unphysical regions where the heavy neutrino width exceeds 30\% of its mass ($\Gamma_N/M_N > 0.3$) are shaded in dark grey, while intermediate-width regions ($0.1 < \Gamma_N/M_N < 0.3$) are shown in light pink. \label{fig:bdt}}
\end{figure*}

Next we examine the relative contribution of each input feature to the BDT's classification power as quantified by the gain-based importance metric. As an example, figure~\ref{fig:importance} shows results for $\mu^+\mu^-$ collisions at $\sqrt{s}=10$~TeV (left) and $\mu^+\mu^+$ collisions at $\sqrt{s}=2$~TeV (right). Although many kinematic variables were provided to the model, only a few emerge as truly impactful. These features are found to be mostly uncorrelated, which allows the BDT to construct independent decision boundaries and avoid redundancy in the classification process. In the case of a 10~TeV opposite-sign muon collider, the azimuthal separation between the leading muons $\Delta\phi(\mu_1, \mu_2)$ stands out as the most discriminating variable, followed by the invariant mass $M_{\mu_1\mu_2}$ and the scalar sum of the muons transverse momenta $p_T^{\mu_1} + p_T^{\mu_2}$. Their ratio $p_T^{\mu_1}/p_T^{\mu_2}$, as well as their recoil $M_\mathrm{recoil}$ contribute next, while still significantly.  This ranking reflects the model's ability to capture the complementary angular and energy dependence of the signal, especially in the VBF-dominated regime. In contrast, pseudo-rapidity-based observables, critical in our cut-based approach to achieve acceptable background rejection, appear to contribute little to the multivariate classification. This can be attributed to correlations with other variables that make them redundant in the presence of more directly informative features. In contrast, for the 2~TeV same-sign  muon collider, the same variables for the final-state muon pseudo-rapidities play a much more important role, the signal being here purely due to VBF-induced contributions, with the subleading variables being similar as in the opposite-sign case.

We remember that while including correlated variables may help improve raw sensitivity, they can complicate the estimation of systematic uncertainties as correlations are more difficult to propagate. In our case, however, only a limited subset of largely uncorrelated input features significantly contributes to the discriminative power of the model. By efficiently exploiting orthogonal information across the input space, while down-weighting redundant variables, our BDT models can then be considered reliable for signal and background classification.

We now apply the two previously trained BDT models to derive projected exclusion limits in the $(M_N, |V_{\mu N}|^2)$ plane. After optimising the selection thresholds for each point in this plane to maximise sensitivity, the exclusion contours are extracted following the procedure outlined in section~\ref{sec:cuts}. The results are presented in figure~\ref{fig:bdt}, where we compare the reach of $\mu^+\mu^-$ collisions at $\sqrt{s} = 10$~TeV with a luminosity of 10~ab$^{-1}$ (blue) to that of $\mu^+\mu^+$ collisions at $\sqrt{s} = 2$~TeV with 1~ab$^{-1}$ (olive). Compared to the cut-based approach (see figure~\ref{fig:res_cutcount}), the BDT analysis leads to a substantial sensitivity gain, particularly in the heavy mass regime with $M_N \gtrsim 10$~TeV and for a collider operating at $\sqrt{s}=10$~TeV. In this region, the signal is dominated by VBF-induced production, which the BDT is especially well-suited to identify thanks to its ability to exploit multidimensional correlations in the final-state kinematics. In particular, Figure~\ref{fig:importance} shows that the non-linear correlations among the $\Delta\phi(\mu_1,\mu_2)$, $M_{\mu_1\mu_2}$ and $p_T^{\mu_1} + p_T^{\mu_2}$ variables play a key role in enhancing the sensitivity, those correlations being otherwise inaccessible through sequential cuts. Moreover, the fixed selection criteria used in the cut-based analysis were optimised for specific benchmark scenarios, and thus become increasingly suboptimal at higher $M_N$ values where the signal kinematics deviate significantly from those benchmarks. The BDT analysis hence extends the reach down to $|V_{\mu N}|^2 \sim 10^{-6}$ for the largest masses considered, improving the results over those of cut-based limits by about two orders of magnitude. 

\section{Summary and Conclusions}\label{sec:conclusions}

In this study, we explored the sensitivity of future muon colliders to heavy Majorana neutrinos, focusing on both opposite-sign ($\mu^+\mu^-$) and same-sign ($\mu^+\mu^+$) configurations. We considered realistic collider scenarios, including centre-of-mass energies of 1, 3 and 10~TeV for $\mu^+\mu^-$ machines and 2~TeV for the proposed $\mu$TRISTAN facility. Our analysis then leveraged VBF processes in the $t$-channel as the dominant production mechanism of Majorana neutrinos. Using both cut-based and multivariate (BDT-based) analysis techniques, we assessed the exclusion potential in the $(M_N, |V_{\mu N}|^2)$ plane on the basis of state-of-the-art simulations, made possible through the development and validation of an SFS muon collider detector card for the \textsc{MadAnalysis5} framework.

We found that future muon colliders could probe heavy neutrinos with masses up to 100~TeV, well beyond the reach of current experiments, and active-sterile mixings as small as $|V_{\mu N}|^2 \sim 10^{-6}$ in the most favourable benchmark scenarios. Furthermore, our results highlight a strong complementarity between collider types: while same-sign $\mu^+\mu^+$ collisions isolate the $t$-channel contribution cleanly, opposite-sign $\mu^+\mu^-$ setups benefit from both $s$-channel and $t$-channel interference and a broader kinematic phase space. In particular, our BDT analysis proves powerful at high energies, fully exploiting the angular and energy correlations induced by heavy Majorana $t$-channel exchange. 

Altogether, our findings underscore the unique discovery potential of high-energy opposite-sign and same-sign muon colliders to probe the Majorana nature of neutrinos across a vast mass range, extending the scope of earlier studies available from the literature by targeting the super-heavy mass regime. This therefore provides strong theoretical motivation for their continued development, with the prospect of shedding light on one of the most pressing open questions in particle physics, the origins of neutrino masses.

\acknowledgments
The authors are grateful to Arindam~Das for enlightening comments on the manuscript. The work of MF has been partly supported by NSERC through grant number SAP105354, and that of BF by the grant ANR-21-CE31-0013 (project DMwithLLPatLHC) from the French \emph{Agence Nationale de la Recherche}. For the purpose of open access, a CC-BY public copyright license has been applied by the authors to the present document, and will be applied to all subsequent versions up to the one accepted for a publication arising from this submission.

\bibliography{ref}
\end{document}